\documentclass[preprint,eqsecnum,aps,nofootinbib]{revtex4}
\usepackage{amsfonts,amsmath,amssymb,amsthm}
\usepackage{latexsym}
\usepackage{bbm,bm}
\usepackage{graphicx}

\newcommand{\ket}[1]{\lvert #1 \rangle}
\newcommand{\bra}[1]{\langle #1 \lvert}
\newcommand{\beq}{\begin{equation}}
\newcommand{\eeq}{\end{equation}}
\newcommand{\beqs}{\begin{eqnarray}}
\newcommand{\eeqs}{\end{eqnarray}}

\begin{document}

\title{Tripartite Entanglement Dynamics in the presence of Markovian or Non-Markovian Environment}

\author{ DaeKil Park$^{1,2}$\footnote{dkpark@kyungnam.ac.kr}}

\affiliation{$^1$Department of Electronic Engineering, Kyungnam University, Changwon
                 631-701, Korea    \\
             $^2$Department of Physics, Kyungnam University, Changwon
                  631-701, Korea    
                      }

\begin{abstract}
We study on the tripartite entanglement dynamics when each party is initially entangled with other parties, but they locally interact with their own Markovian
or  non-Markovian environment. First, we consider three GHZ-type initial states, all of which have GHZ symmetry provided that the parameters are 
chosen appropriately. However, this symmetry is broken due to the effect of environment. The corresponding  $\pi$-tangles, 
one of the tripartite entanglement measure, are analytically computed at arbitrary time. For Markovian case while the tripartite entanglement for type I exhibits
an entanglement sudden death, the dynamics for the remaining cases decays normally in time with the half-life rule.
For non-Markovian case the revival phenomenon of entanglement occurs after 
complete disappearance of entanglement. We also consider two W-type initial states. For both cases  the $\pi$-tangles are analytically derived.
The revival phenomenon also occurs in this case. On the analytical ground the robustness or fragility issue against the effect of environment is examined 
for both GHZ-type and W-type initial states.
\end{abstract}

\maketitle
\section{Introduction}
Entanglement\cite{nielsen00,horodecki09} is one of the important concepts from fundamental aspect of quantum mechanics and practical aspect of quantum information processing. As shown for last two decades it plays a crucial role in quantum teleportation\cite{teleportation},
superdense coding\cite{superdense}, quantum cloning\cite{clon}, and quantum cryptography\cite{cryptography,cryptography2}. It is also quantum entanglement, 
which makes the quantum computer\footnote{The current status of quantum computer technology was reviewed in Ref.\cite{qcreview}.} outperform the classical one\cite{qcomputer}.

Quantum mechanics is a physics, which is valid for ideally closed system. However, real physical systems inevitably interact with their 
surroundings. Thus, it is important to study how the environment modifies the dynamics of given physical system. There are two different tools
for describing the evolution of open quantum system: quantum operation formalism\cite{nielsen00} and master equation approach\cite{breuer02}.
Both tools have their own merits.

Since it is known that quantum system loses quantum properties by contacting the environment\cite{zurek03}, we expect that the degradation 
of entanglement occurs\cite{yu02-1,simon02-1,dur04-1}. Sometimes entanglement exhibits an exponential decay in time by successive halves. 
Sometimes, however, the entanglement sudden death (ESD) occurs when the entangled multipartite quantum system is embedded in
Markovian environments\cite{markovian,yu05-1,yu06-1,yu09-1}. This means that the entanglement is completely disentangled at finite times. 
This ESD phenomenon has been revealed experimentally\cite{almeida07,laurat07}. When the ESD occurs, it is natural to ask where the lost 
entanglement goes. It was found that when the entanglement of given quantum system suddenly disappears, the reservoir entanglement suddenly 
appears, which is called entanglement sudden birth (ESB) \cite{lopez08}. Since we do not consider the degrees of freedom for the environment, we do not examine the ESB phenomenon in this paper.

The dynamics of entanglement was also examined when the physical system is embedded in non-Markovian environment\cite{breuer02,bellomo07}.
It has been shown that there is a revival of entanglement after a finite period of time of its complete disappearance. This is mainly due to 
the memory effect of the non-Markovian environment. This phenomenon was shown in Ref.\cite{bellomo07} by making use 
of the two qubit system and concurrence\cite{concurrence1} as a bipartite entanglement measure. Subsequently, many works have been done 
to quantify the non-Markovianity\cite{breuer09,vacchini11,chruscinski11,rivas14,hall14,kwang15-1}.

In this paper we consider the  entanglement dynamics when the qubit system interacts with the Markovian or non-Markovian environment. 
So far this issue was investigated by making use of the bipartite system. Recently, the tripartite entanglement dynamics was also explored in 
Ref.\cite{kwang15-1} numerically. Since entanglement is an important physical resource in the quantum information processing, it is important
to control the entanglement dynamics when the environment is present. 
In order to control the entanglement it is crucial to derive the entanglement analytically in the 
entire range of time. For example, analytic derivation for the bipartite entanglement dynamics enables us to explore the entanglement invariants\cite{yonac07,yu09-1}.
It is also possible to discuss the robustness or fragility issue against the environment by exploiting the analytical results.
Thus, we will explore the tripartite entanglement dynamics in this paper on the analytical ground. For simplicity, we
choose the physical setting, i.e. there is no interaction between qubit and each qubit interacts with its own reservoir. We will compute the 
entanglement at arbitrary time for three-types of initial Greenberger-Horne-Zeilinger(GHZ) states\cite{green89} and for 
two types of initial W-states\cite{dur00-1} in the presence of the Markovian or non-Markovian environment.

Typical tripartite entanglement measures are residual entanglement\cite{ckw} and $\pi$-tangle\cite{ou07-1}.
For three-qubit pure state
$|\psi\rangle = \sum_{i,j,k=0}^1 a_{ijk} |ijk\rangle$ the residual entanglement $\tau_{ABC}$
becomes
\begin{equation}
\label{residual-1}
\tau_{ABC} = 4 |d_1 - 2 d_2 + 4 d_3|,
\end{equation}
where
\begin{eqnarray}
\label{residual-2}
& &d_1 = a^2_{000} a^2_{111} + a^2_{001} a^2_{110} + a^2_{010} a^2_{101} +
                                                              a^2_{100} a^2_{011
},
                                                              \\   \nonumber
& &d_2 = a_{000} a_{111} a_{011} a_{100} + a_{000} a_{111} a_{101} a_{010} +
         a_{000} a_{111} a_{110} a_{001}
                                                              \\   \nonumber
& &\hspace{1.0cm} +
         a_{011} a_{100} a_{101} a_{010} + a_{011} a_{100} a_{110} a_{001} +
         a_{101} a_{010} a_{110} a_{001},
                                                              \\   \nonumber
& &d_3 = a_{000} a_{110} a_{101} a_{011} + a_{111} a_{001} a_{010} a_{100}.
\end{eqnarray}
Thus, the residual entanglement of any three-qubit pure state can be computed by making use of Eq. (\ref{residual-1}). Although the residual entanglement
can detect the GHZ-type entanglement, it cannot detect the W-type entanglement:
\begin{equation}
\label{residual-3}
\tau_{ABC} (GHZ) = 1 \hspace{1.0cm} \tau_{ABC} (W) = 0,
\end{equation}
where
\begin{equation}
\label{residual-4}
\ket{GHZ} = \frac{1}{\sqrt{2}} \left( \ket{000} + \ket{111} \right)
\hspace{1.0cm} \ket{W} = \frac{1}{\sqrt{3}} \left( \ket{001} + \ket{010} + \ket{100} \right).
\end{equation}

For mixed states the residual entanglement is defined by a convex-roof
method\cite{benn96,uhlmann99-1} as follows:
\begin{equation}
\label{residual-5}
\tau_{ABC} (\rho) = \min \sum_i p_i \tau_{ABC} (\rho_i),
\end{equation}
where the minimum is taken over all possible ensembles of pure states. The pure state ensemble
corresponding to the minimum $\tau_{ABC}$ is called the optimal decomposition. It is in general
difficult to derive the optimal decomposition for arbitrary mixed states. Hence, analytic
computation of the residual entanglement can be done for rare cases\cite{residual}.
 Furthermore, recently, 
three-tangle\footnote{In this paper we will call $\tau_3 = \sqrt{\tau_{ABC}}$ three-tangle and $\tau_3^2 = \tau_{ABC}$ 
residual entanglement.} $\tau_3$ of the whole GHZ-symmetric states\cite{elts12-1} was explicitly computed\cite{siewert12-1}.

The $\pi$-tangle defined in Ref.\cite{ou07-1} is easier for analytic computation than the residual entanglement (or three tangle) because it 
does not rely on the convex-roof method. 
 The $\pi$-tangle is
defined in terms of the global negativities~\cite{vidal01-1}. For a three-qubit state $\rho$ 
they are given by
\begin{equation}
\label{negativity-1}
{\cal N}^A = || \rho^{T_A} || - 1, \hspace{1.0cm}
{\cal N}^B = || \rho^{T_B} || - 1, \hspace{1.0cm}
{\cal N}^C = || \rho^{T_C} || - 1,
\end{equation}
where $||R|| = \mbox{Tr} \sqrt{R R^{\dagger}}$, and the superscripts $T_A$, $T_B$, and $T_C$
represent the partial transposes of $\rho$ with respect to the qubits $A$, $B$, and $C$ respectively. 
Then, the $\pi$-tangle is defined as
\begin{equation}
\label{pi-1}
\pi_{ABC} = \frac{1}{3} (\pi_A + \pi_B + \pi_C ),
\end{equation}
where
\begin{equation}
\label{pi-2}
\pi_A = {\cal N}_{A(BC)}^2 - ({\cal N}_{AB}^2 + {\cal N}_{AC}^2) \hspace{.5cm}
\pi_B = {\cal N}_{B(AC)}^2 - ({\cal N}_{AB}^2 + {\cal N}_{BC}^2) \hspace{.5cm}
\pi_C = {\cal N}_{(AB)C}^2 - ({\cal N}_{AC}^2 + {\cal N}_{BC}^2).
\end{equation}
The remarkable property of the $\pi$-tangle is that it can detect not only the GHZ-type entanglement but also the W-type
entanglement:
\begin{equation}
\label{pi-ghz-w}
\pi_{ABC} (GHZ) = 1 \hspace{1.0cm}
\pi_{ABC} (W) = \frac{4}{9} (\sqrt{5} - 1) \sim 0.55.
\end{equation}

As commented earlier we will examine the tripartite entanglement dynamics of the three-qubit states in the presence of the Markovian or non-Markovian
environment. We will adopt the $\pi$-tangle as an entanglement measure for analytic computation as much as possible. In section II we consider 
how the three-qubit initial state is evolved when each qubit interacts with its own Markovian or non-Markovian environment\cite{bellomo07}. In section III we 
explore the entanglement dynamics of three GHZ-type initial states. The initial states are local unitary (LU) with each other. Thus, their entanglement are 
the same initially. Furthermore, if the parameters are appropriately chosen, they all have GHZ-symmetry, i.e. they are invariant under 
(i) qubit permutation (ii) simultaneous three-qubit flips (iii) qubit rotations about the $z$-axis. However, this symmetry is broken due to the 
environment effect. As a result, their entanglement dynamics are different with each other. In section IV we examine the entanglement 
dynamics of two W-type initial states. They are also LU to each other. However, the dynamics is also different because of the environment effect. 
In section V a brief conclusion is given.

\section{General Features}
We consider three-qubit system, each of which interacts only and independently with its local environment. We assume that the dynamics of 
single qubit is governed by Hamiltonian
\begin{equation}
\label{hamitonian-1}
H = H_0 + H_I
\end{equation}
where
\begin{eqnarray}
\label{hamiltonian-2}
& &H_0 = \omega_0 \sigma_+ \sigma_- + \sum_{k} \omega_k b_k^{\dagger} b_k                      \\      \nonumber
& &H_I = \sigma_+ \otimes B + \sigma_- \otimes B^{\dagger} \hspace{1.0cm} \mbox {with}\hspace{.5cm} B = \sum_k g_k b_k.
\end{eqnarray} 
In Eq. (\ref{hamiltonian-2}) $\omega_0$ is a transition frequency of the two-level system (qubit), and $\sigma_{\pm}$ are the raising and 
lowering operators. The index $k$ labels the different field modes of the reservoir with frequencies $\omega_k$, creation and annihilation 
operators $ b_k^{\dagger}$, $b_k$, and coupling constants $g_k$.  In the interaction picture the dynamics is governed by the Schr\"{o}dinger
equation
\begin{equation}
\label{schrodinger-1}
\frac{d}{d t} \psi (t) = -i H_I (t) \psi(t)
\end{equation}
where
\begin{eqnarray}
\label{schrodinger-2}
&&H_I (t) \equiv e^{i H_0 t} H_I e^{-i H_0 t} = \sigma_+ (t)\otimes  B(t) + \sigma_- (t) \otimes B^{\dagger} (t)   \nonumber  \\
&&\sigma_{\pm} (t) \equiv  e^{i H_0 t} \sigma_{\pm} e^{-i H_0 t} = \sigma_{\pm} e^{\pm i \omega_0 t}    \\   \nonumber
&&B(t) \equiv e^{i H_0 t} B e^{-i H_0 t} = \sum_k g_k b_k e^{-i \omega_k t}.
\end{eqnarray}
The Hamiltonian (\ref{hamitonian-1}) represents one of few exactly solvable model\cite{garraway97}. This means that the Schr\"{o}dinger
equation (\ref{schrodinger-1}) can be formally solved if $\psi (0)$ is given. Then, the reduced state of the single qubit 
$\hat{\rho}^S (t) \equiv Tr_{env} \ket{\psi(t)} \bra{\psi(t)}$ is given by\cite{breuer02,manis06}
\begin{eqnarray}
\label{density-1}
\hat{\rho}^S (t) = \left(    \begin{array}{cc}
                                    \rho_{00}^S (0) + \rho_{11}^S (0) \left( 1 - |P_t|^2 \right) &   \rho_{01}^S (0)  P_t     \\
                                    \rho_{10}^S (0) P_t^*    &   \rho_{11}^S (0)  |P_t|^2
                                         \end{array}                                                       \right)
\end{eqnarray}
where $\hat{\rho}^S (0) =  Tr_{env} \ket{\psi(0)} \bra{\psi(0)}$ and $Tr_{env}$ denotes the partial trace over the environment.
The function $P_t$ satisfies the differential equation
\begin{equation}
\label{density-2}
\frac{d}{dt} P_t = - \int_0^t dt_1 f(t - t_1) P_{t_1}
\end{equation}
and the correlation function $f(t - t_1)$ is related to the spectral density $J(\omega)$ of the reservoir by
\begin{equation}
\label{density-3}
f(t - t_1) = \int J(\omega) exp[i (\omega_0 - \omega) (t - t_1)].
\end{equation}
We choose $J(\omega)$ as an effective spectral density of the damped Jaynes-Cummings model\cite{breuer02}
\begin{equation}
\label{jc-1}
J(\omega) = \frac{1}{2 \pi} \frac{\gamma_0 \lambda^2}{(\omega_0 - \omega)^2 + \lambda^2}
\end{equation}
where the parameter $\lambda$ defines the spectral width of the coupling, which is connected to the reservoir 
correlation time $\tau_B$ by the relation $\tau_B = 1 / \lambda$ and the relaxation time scale $\tau_R$ on which the state of the system
changes is related to $\gamma_0$ by $\tau_R = 1 / \gamma_0$.

By making use of the Residue theorem in complex plane the correlation function can be easily computed in a form
\begin{equation}
\label{correlation-1}
f(t - t_1) = \frac{\gamma_0 \lambda}{2} e^{-\lambda |t - t_1|}.
\end{equation}
Inserting Eq. (\ref{correlation-1}) into Eq. (\ref{density-2}) and making use of Laplace transform, one can compute $P_t$ explicitly. While in a 
weak coupling (or Markovian) regime $\tau_R > 2 \tau_B$ $P_t$ becomes
\begin{equation}
\label{pt-m}
P_t = e^{-\frac{\lambda}{2} t} \left[ \cosh \left( \frac{\bar{d}}{2} t \right) + \frac{\lambda}{\bar{d}} \sinh \left( \frac{\bar{d}}{2} t \right)
                                              \right]
\end{equation}
with $\bar{d} = \sqrt{\lambda^2 - 2 \gamma_0 \lambda}$, in a strong coupling (or non-Markovian) regime $\tau_R < 2 \tau_B$  $P_t$ reduces to 
\begin{equation}
\label{pt-nm}
P_t = e^{-\frac{\lambda}{2} t} \left[ \cos \left( \frac{d}{2} t \right) + \frac{\lambda}{d} \sin \left( \frac{d}{2} t \right)
                                              \right]
\end{equation}
with $d = \sqrt{2 \gamma_0 \lambda - \lambda^2}$. Since, in the Markovian regime $\lambda > 2 \gamma_0$, $P_t$ in Eq. (\ref{pt-m}) exhibits
an exponential decay in time, it seems to make exponential decay of entanglement  or ESD phenomenon. 
However, in the non-Markovian regime  $\lambda < 2 \gamma_0$, $P_t$ in 
Eq. (\ref{pt-nm}) exhibits an oscillatory behavior in time with decreasing amplitude. It seems to be responsible for the revival phenomenon of 
entanglement\cite{bellomo07}, after a finite period of time of its complete disappearance.

The state $\hat{\rho}^T (t)$ at time $t$ of whole three-qubit system, each of which interacts only and independently with its own environment, can be derived by the Kraus operators\cite{kraus83}. Introducing, for simplicity, $\{ \ket{0} \equiv \ket{000},  \ket{1} \equiv \ket{001}, 
 \ket{2} \equiv \ket{010},  \ket{3} \equiv \ket{011},  \ket{4} \equiv \ket{100}, \ket{5} \equiv \ket{101},  \ket{6} \equiv \ket{110}, 
 \ket{7} \equiv \ket{111} \}$, the diagonal parts of $\hat{\rho}^T (t)$ are
 \begin{eqnarray}
 \label{diagonal-1}
 &&\hat{\rho}^T_{11} (t) = P_t^2 \left[\hat{\rho}^T_{11} (0) + \left\{ \hat{\rho}^T_{33} (0) + \hat{\rho}^T_{55} (0) \right\} (1 - P_t^2) + 
 \hat{\rho}^T_{77} (0) (1 - P_t^2)^2 \right]                                                       \nonumber   \\
 &&\hat{\rho}^T_{22} (t) = P_t^2 \left[\hat{\rho}^T_{22} (0) + \left\{ \hat{\rho}^T_{33} (0) + \hat{\rho}^T_{66} (0) \right\} (1 - P_t^2) + 
 \hat{\rho}^T_{77} (0) (1 - P_t^2)^2 \right]                                                       \nonumber   \\
 &&\hat{\rho}^T_{33} (t) = P_t^4 \left[\hat{\rho}^T_{33} (0) + \hat{\rho}^T_{77} (0) (1 - P_t^2) \right]     \\    \nonumber
  &&\hat{\rho}^T_{44} (t) = P_t^2 \left[\hat{\rho}^T_{44} (0) + \left\{ \hat{\rho}^T_{55} (0) + \hat{\rho}^T_{66} (0) \right\} (1 - P_t^2) + 
 \hat{\rho}^T_{77} (0) (1 - P_t^2)^2 \right]                                                                                                \\    \nonumber
 &&\hat{\rho}^T_{55} (t) = P_t^4 \left[\hat{\rho}^T_{55} (0) + \hat{\rho}^T_{77} (0) (1 - P_t^2) \right]     \\    \nonumber      
 &&\hat{\rho}^T_{66} (t) = P_t^4 \left[\hat{\rho}^T_{66} (0) + \hat{\rho}^T_{77} (0) (1 - P_t^2) \right]     \\    \nonumber   
 &&\hat{\rho}^T_{00} (t) = 1 - \sum_{i=1}^7 \hat{\rho}^T_{ii} (t)
 \end{eqnarray}
and the non-diagonal parts are 
\begin{eqnarray}
\label{non-diagonal-1}
&&\hat{\rho}^T_{01} (t) = P_t \left[\hat{\rho}^T_{01} (0) + \left\{ \hat{\rho}^T_{23} (0) + \hat{\rho}^T_{45} (0) \right\} (1 - P_t^2) + 
 \hat{\rho}^T_{67} (0) (1 - P_t^2)^2 \right]         \nonumber   \\
&&\hat{\rho}^T_{02} (t) = P_t \left[\hat{\rho}^T_{02} (0) + \left\{ \hat{\rho}^T_{13} (0) + \hat{\rho}^T_{46} (0) \right\} (1 - P_t^2) + 
 \hat{\rho}^T_{57} (0) (1 - P_t^2)^2 \right]         \nonumber   \\
&&\hat{\rho}^T_{04} (t) = P_t \left[\hat{\rho}^T_{04} (0) + \left\{ \hat{\rho}^T_{15} (0) + \hat{\rho}^T_{26} (0) \right\} (1 - P_t^2) + 
 \hat{\rho}^T_{37} (0) (1 - P_t^2)^2 \right]         \nonumber   \\
&&\hat{\rho}^T_{03} (t) = P_t^2 \left[\hat{\rho}^T_{03} (0) + \hat{\rho}^T_{47} (0) (1 - P_t^2) \right]  \hspace{.5cm}
\hat{\rho}^T_{05} (t) = P_t^2 \left[\hat{\rho}^T_{05} (0) + \hat{\rho}^T_{27} (0) (1 - P_t^2) \right]      \nonumber   \\
&&\hat{\rho}^T_{06} (t) = P_t^2 \left[\hat{\rho}^T_{06} (0) + \hat{\rho}^T_{17} (0) (1 - P_t^2) \right]  \hspace{.5cm}
\hat{\rho}^T_{12} (t) = P_t^2 \left[\hat{\rho}^T_{12} (0) + \hat{\rho}^T_{56} (0) (1 - P_t^2) \right]      \nonumber   \\
&&\hat{\rho}^T_{13} (t) = P_t^3 \left[\hat{\rho}^T_{13} (0) + \hat{\rho}^T_{57} (0) (1 - P_t^2) \right]  \hspace{.5cm}
\hat{\rho}^T_{14} (t) = P_t^2 \left[\hat{\rho}^T_{14} (0) + \hat{\rho}^T_{36} (0) (1 - P_t^2) \right]      \nonumber   \\
&&\hat{\rho}^T_{15} (t) = P_t^3 \left[\hat{\rho}^T_{15} (0) + \hat{\rho}^T_{37} (0) (1 - P_t^2) \right]  \hspace{.5cm}
\hat{\rho}^T_{23} (t) = P_t^3 \left[\hat{\rho}^T_{23} (0) + \hat{\rho}^T_{67} (0) (1 - P_t^2) \right]      \\  \nonumber  
&&\hat{\rho}^T_{24} (t) = P_t^2 \left[\hat{\rho}^T_{24} (0) + \hat{\rho}^T_{35} (0) (1 - P_t^2) \right]  \hspace{.5cm}
\hat{\rho}^T_{26} (t) = P_t^3 \left[\hat{\rho}^T_{26} (0) + \hat{\rho}^T_{37} (0) (1 - P_t^2) \right]      \\  \nonumber   
&&\hat{\rho}^T_{45} (t) = P_t^3 \left[\hat{\rho}^T_{45} (0) + \hat{\rho}^T_{67} (0) (1 - P_t^2) \right]  \hspace{.5cm}
\hat{\rho}^T_{46} (t) = P_t^3 \left[\hat{\rho}^T_{46} (0) + \hat{\rho}^T_{57} (0) (1 - P_t^2) \right]      \\  \nonumber
&&\hat{\rho}^T_{07} (t) = \hat{\rho}^T_{07} (0) P_t^3  \hspace{.5cm} \hat{\rho}^T_{16} (t) = \hat{\rho}^T_{16} (0) P_t^3  \hspace{.5cm}
\hat{\rho}^T_{17} (t) = \hat{\rho}^T_{17} (0) P_t^4  \hspace{.5cm} \hat{\rho}^T_{25} (t) = \hat{\rho}^T_{25} (0) P_t^3   \\  \nonumber
&&\hat{\rho}^T_{27} (t) = \hat{\rho}^T_{27} (0) P_t^4  \hspace{.5cm} \hat{\rho}^T_{34} (t) = \hat{\rho}^T_{34} (0) P_t^3  \hspace{.5cm}
\hat{\rho}^T_{35} (t) = \hat{\rho}^T_{35} (0) P_t^4  \hspace{.5cm} \hat{\rho}^T_{36} (t) = \hat{\rho}^T_{36} (0) P_t^4   \\  \nonumber
&&\hspace{1.3cm}\hat{\rho}^T_{37} (t) = \hat{\rho}^T_{37} (0) P_t^5  \hspace{.5cm} \hat{\rho}^T_{47} (t) = \hat{\rho}^T_{47} (0) P_t^4  \hspace{.5cm} \hat{\rho}^T_{56} (t) = \hat{\rho}^T_{56} (0) P_t^4     \\  \nonumber
&&\hspace{2.7cm}\hat{\rho}^T_{57} (t) = \hat{\rho}^T_{57} (0) P_t^5  \hspace{.5cm} \hat{\rho}^T_{67} (t) = \hat{\rho}^T_{67} (0) P_t^5
\end{eqnarray}
with $\hat{\rho}^T_{ij} (t) = \hat{\rho}^{T*}_{ji} (t)$. Now, we are ready to explore the tripartite entanglement dynamics in the presence of the 
Markovian or non-Markovian environment.

\section{entanglement dynamics of GHZ-type initial states}
In this section we examine the tripartite entanglement dynamics when the initial states are GHZ-type states. All initial states have 
GHZ-symmetry\cite{elts12-1} if the parameters are appropriately chosen. However, this symmetry is broken due to the effects of environment.

\subsection{Type I}
Let us choose the initial state in a form
\begin{equation}
\label{type1-ghz-1}
\hat{\rho}^T_I (0) = \ket{\psi_I} \bra{\psi_I}
\end{equation}
where
$\ket{\psi_I} = a \ket{0} + b e^{i \delta} \ket{7}$ with $a^2 + b^2 = 1$. As commented before $\ket{\psi_I}$ has a GHZ-symmetry when 
$a^2 = b^2 = 1/2$ and $\delta = 0$. Then the spectral decomposition of   $\hat{\rho}^T_I (t)$ can be read directly from Eqs. (\ref{diagonal-1}) 
and (\ref{non-diagonal-1}) as a form:
\begin{eqnarray}
\label{type1-ghz-2}
&&\hat{\rho}^T_I (t) = \Lambda_+ \ket{\psi_1} \bra{\psi_1} + \Lambda_-   \ket{\psi_2} \bra{\psi_2} 
+  b^2 P_t^2 (1 - P_t^2)^2 \left\{ \ket{1} \bra{1} +   \ket{2} \bra{2} +  \ket{4} \bra{4} \right\}                   \\    \nonumber
&&\hspace{3.0cm} + b^2 P_t^4 (1 - P_t^2) \left\{  \ket{3} \bra{3} +  \ket{5} \bra{5} +  \ket{6} \bra{6} \right\}
\end{eqnarray}
where
\begin{equation}
\label{type1-ghz-3}
\Lambda_{\pm} = \frac{1}{2} \left[ \left\{ 1 - 3 b^2 P_t^2 (1 - P_t^2) \right\} 
\pm \sqrt{\left[  1 - 3 b^2 P_t^2 (1 - P_t^2) \right]^2 - 4 b^4 P_t^6 (1 - P_t^2)^2 } \right]
\end{equation}
and 
\begin{equation}
\label{type1-ghz-4}
\ket{\psi_1} = \frac{1}{N_I} \left( x \ket{0} + y e^{i \delta} \ket{7} \right)   \hspace{1cm}
\ket{\psi_2} = \frac{1}{N_I} \left( y \ket{0} - x e^{i \delta} \ket{7} \right)
\end{equation}
with 
\begin{eqnarray}
\label{type1-ghz-5}
&&x = 1 - b^2 P_t^2 (3 - 3 P_t^2 + 2 P_t^4) +  \sqrt{\left[  1 - 3 b^2 P_t^2 (1 - P_t^2) \right]^2 - 4 b^4 P_t^6 (1 - P_t^2)^2 }   \nonumber \\
&&y = 2 a b P_t^2
\end{eqnarray}
and $N_I = \sqrt{x^2 + y^2}$ is a normalization constant.

Since $\hat{\rho}^T_I (t)$ is a full rank, it seems to be highly difficult to compute the residual entanglement (or three-tangle) analytically. However,
from Eq. (\ref{type1-ghz-2}) one can realize the upper bound of $\tau_{ABC}$ as 
\begin{equation}
\label{upper-1}
\tau_{ABC} \leq \left[1 - 3 b^2 P_t^2 (1 - P_t^2)\right] \frac{4 x^2 y^2}{(x^2 + y^2)^2}.
\end{equation}
It is worthwhile noting that  $\hat{\rho}^T_I (t)$ does not have the GHZ-symmetry even at $a^2 = b^2 = 1/2$ and $\delta = 0$. Thus, the symmetry 
which  $\hat{\rho}^T_I (0)$ has is broken due to the effect of environment.

In order to explore the tripartite entanglement dynamics on the analytical ground, we compute the $\pi$-tangle of $\hat{\rho}^T_I (t)$. Using 
Eq. (\ref{negativity-1}) it is straightforward to compute the induced bipartite entanglement quantities 
${\cal N}_{A(BC)}$,  ${\cal N}_{B(AC)}$, and  ${\cal N}_{(AB)C}$. One can show that they are all the same with
\begin{eqnarray}
\label{ghz-1-pi-1}
{\cal N}_{A(BC)} = {\cal N}_{B(AC)} = {\cal N}_{(AB)C} = \max \left[ Q(t), 0 \right],
\end{eqnarray}
where
\begin{equation}
\label{add1}
Q(t) = \sqrt{b^4 P_t^4 (1 - P_t^2)^2 (1 - 2 P_t^2)^2 + 4 a^2 b^2 P_t^6} - b^2 P_t^2 (1 - P_t^2).
\end{equation}
One can also show the two-tangles completely vanish, i.e. ${\cal N}_{AB} = {\cal N}_{AC} = {\cal N}_{BC} = 0$, easily. Thus the $\pi$-tangle of  $\hat{\rho}^T_I (t)$ is 
\begin{equation}
\label{ghz-1-pi-2}
\pi^I_{GHZ} (t) = {\cal N}_{A(BC)}^2.
\end{equation}

\begin{figure}[ht!]
\begin{center}
\includegraphics[height=4.9cm]{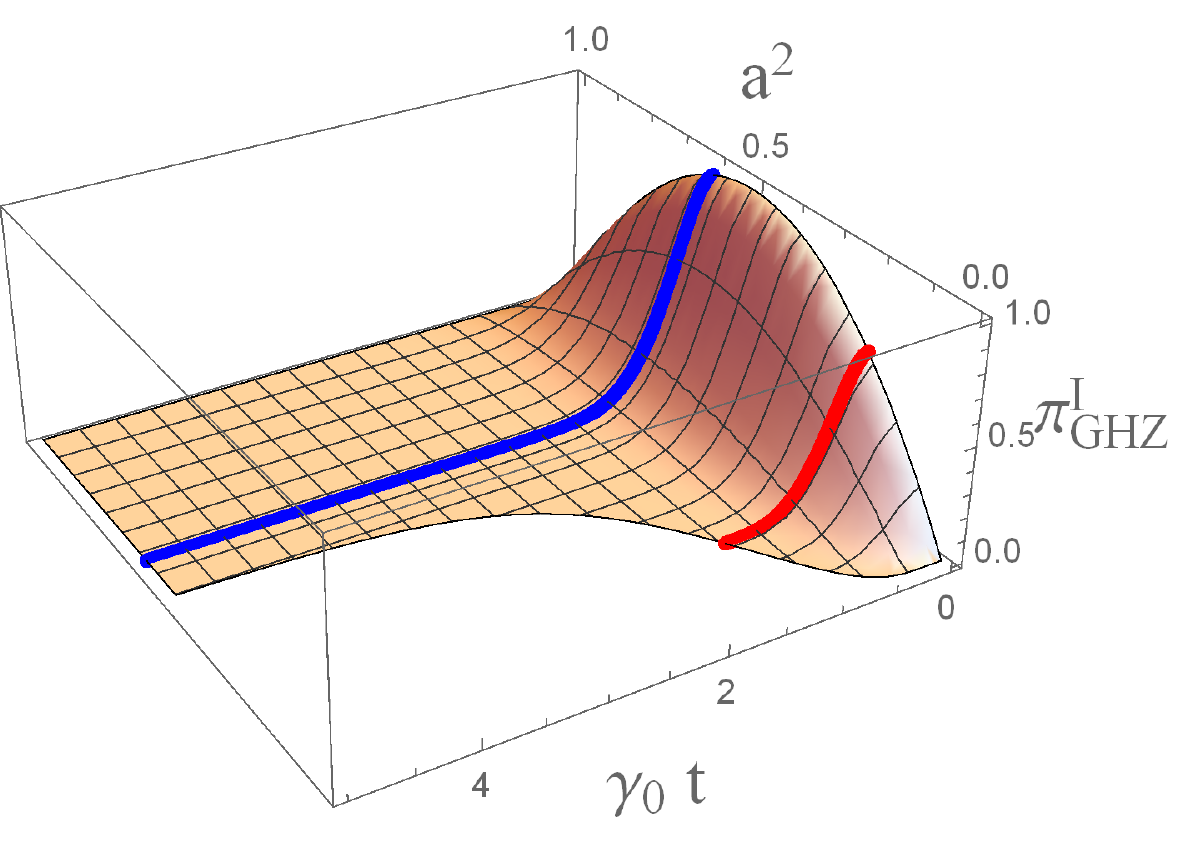}
\includegraphics[height=4.9cm]{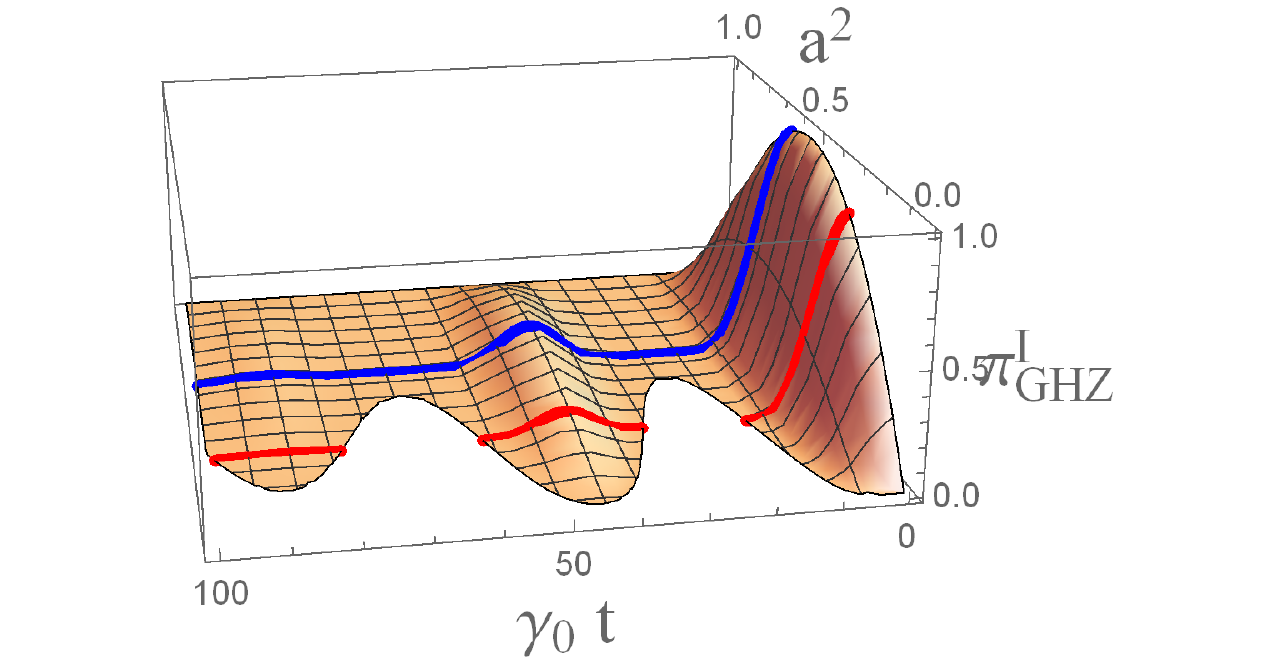}
\caption[fig1]{(Color online) The $\pi$-tangle of  $\hat{\rho}^T_I (t)$ as a function of the parameters $\gamma_0 t$ 
and $a^2$ when the state  interacts with the Markovian and non-Markovian environments. We choose $\lambda$ as (a)  $\lambda = 3 \gamma_0$ and 
(b) $\lambda = 0.01 \gamma_0$. }
\end{center}
\end{figure}

Eq. (\ref{ghz-1-pi-1}) guarantees that regardless of Markovian or non-Markovian environment $\pi^I_{GHZ} (t)$ becomes zero if an inequality
\begin{equation}
\label{add1}
a^2 \leq \frac{(1 - P_t^2)^3}{1 + (1 - P_t^2)^3}
\end{equation}
holds because $Q(t)$ becomes negative in this condition. 

Now, let us examine the dynamics of the tripartite entanglement for $\hat{\rho}^T_I (t)$ when the quantum system interacts with Markovian environment.
Since $P_t$ in Eq. (\ref{pt-m}) decays exponentially in time, one can expect that the tripartite entanglement exhibits an asymptotic decay, 
i.e. decay with the half-life rule, similarly. In fact, this is true when the inequality (\ref{add1}) is violated. If the inequality holds at $t \geq t_*$, the tripartite entanglement 
becomes zero at $t = t_*$ abruptly. This is an ESD phenomenon. If the inequality does not hold for all time, the tripartite entanglement decays with the 
half-life rule as expected. This is shown clearly in Fig. 1(a), where $\pi^I_{GHZ} (t)$ is plotted as a function of $\gamma_0 t$ and $a^2$. In this 
figure we choose $\lambda = 3 \gamma_0$. As expected, the tripartite entanglement decreases with increasing $\gamma_0 t$. 
When $a^2 = 0.6$ (blue line) it decays exponentially in $\gamma_0 t$ with the half-life rule. For $a^2 = 0.2$ (red line), however, it becomes zero in the region $\gamma_0 t \geq 1.21$.  

For non-Markovian regime the decay behavior of the tripartite entanglement in time is completely different. This difference arises due to  combination 
of the inequality (\ref{add1}) and difference form of $P_t$. Since $P_t$ in Eq. (\ref{pt-nm}) exhibits an underdamping behavior in time with 
zeros  at $t_n = 2[n \pi -  \tan^{-1} (d / \lambda) / d] \hspace{.2cm}  (n = 1, 2, \cdots)$, one may expect that the tripartite entanglement also
decays with oscillatory behavior. This is true when the inequality (\ref{add1}) is violated for all time. This behavior is shown as a blue line 
($a^2 = 0.6$) of Fig. 1(b). In this figure we choose 
$\lambda = 0.01 \gamma_0$. If the inequality holds for some time interval $t_{*1} \leq t \leq t_{*2}$, the tripartite entanglement becomes zero in this 
interval. After this time interval, however, nonzero tripartite entanglement reappears, which makes a revival of entanglement after a finite period of time of 
its complete disappearance. This is shown as a red line ($a^2 = 0.3$) of Fig. 1(b).

\subsection{Type II}
Let us choose the initial state in a form
\begin{equation}
\label{type2-ghz-1}
\hat{\rho}^T_{II} (0) = \ket{\psi_{II}} \bra{\psi_{II}}
\end{equation}
where
$\ket{\psi_{II}} = a \ket{1} + b e^{i \delta} \ket{6}$ with $a^2 + b^2 = 1$. 
Since $\ket{\psi_{I}} = \openone \otimes \openone \otimes \sigma_x \ket{\psi_{II}}$,  $(\openone \otimes \openone \otimes \sigma_x) 
\hat{\rho}^T_{II} (0) (\openone \otimes \openone \otimes \sigma_x)^{\dagger}$ has a GHZ-symmetry provided that $a^2 = b^2 = 1/2$ and 
$\delta = 0$.

Using Eqs. (\ref{diagonal-1}) and (\ref{non-diagonal-1}) one can show that the spectral decomposition of $\hat{\rho}^T_{II} (t)$ becomes
\begin{equation}
\label{type2-ghz-2}
\hat{\rho}^T_{II} (t) = \lambda_2 \ket{\phi_{II}} \bra{\phi_{II}} + (1 - P_t^2) \left[ a^2 + b^2 (1 - P_t^2) \right] \ket{0} \bra{0} + 
b^2 P_t^2 (1 - P_t^2) \left( \ket{2} \bra{2} + \ket{4} \bra{4} \right)
\end{equation}
where
\begin{eqnarray}
\label{type2-ghz-3}
&&\lambda_2 = P_t^2 (a^2 + b^2 P_t^2)                      \\    \nonumber
&& \ket{\phi_{II}} = \frac{1}{\sqrt{a^2 + b^2 P_t^2}} \left( a \ket{1} + b P_t e^{i \delta} \ket{6} \right).
\end{eqnarray}
Unlike the case of type I $\hat{\rho}^T_{II} (t)$ is rank four tensor. From Eq. (\ref{type2-ghz-2}) one can derive the upper bound of $\tau_{ABC}$
for $\hat{\rho}^T_{II} (t)$, which is 
\begin{equation}
\label{upper-2}
\tau_{ABC} \leq \frac{4 a^2 b^2 P_t^4}{a^2 + b^2 P_t^2}.
\end{equation}

The negativities  ${\cal N}_{A(BC)}$,  ${\cal N}_{B(AC)}$, and  ${\cal N}_{(AB)C}$ of  $\hat{\rho}^T_{II} (t)$ can be computed by making use of 
Eq. (\ref{negativity-1}). The final expressions are 
\begin{eqnarray}
\label{ghz-2-pi-1}
&&{\cal N}_{A(BC)} = {\cal N}_{B(AC)} = \sqrt{b^4 P_t^4 (1 - P_t^2)^2 + 4 a^2 b^2 P_t^6} - b^2 P_t^2 (1 - P_t^2)   \\   \nonumber
&&{\cal N}_{(AB)C} = \sqrt{(1 - P_t^2)^2 \left[ a^2 + b^2 (1 - P_t^2) \right]^2 + 4 a^2 b^2 P_t^6} - (1 - P_t^2) \left[ a^2 + b^2 (1 - P_t^2) \right].
\end{eqnarray}
It is also easy to show ${\cal N}_{AB} = {\cal N}_{AC} = {\cal N}_{BC} = 0$. Thus the $\pi$-tangle of $\hat{\rho}^T_{II} (t)$ is 
\begin{equation}
\label{ghz-2-pi-2}
\pi_{GHZ}^{II} (t) = \frac{1}{3} \left[2 {\cal N}_{A(BC)}^2 + {\cal N}_{(AB)C}^2 \right].
\end{equation}
When $t = 0$, $\pi_{GHZ}^{II} (0)$ becomes $4 a^2 b^2$ and it reduces to zero as $t \rightarrow \infty$. Of course, the entanglement of 
 $\hat{\rho}^T_{II} (t)$ is completely disentangled at $t = t_n \hspace{.2cm} (n = 1, 2, \cdots)$ in the non-Markovian regime.
 
 \subsection{Type III}
 Let us choose the initial state in a form
\begin{equation}
\label{type3-ghz-1}
\hat{\rho}^T_{III} (0) = \ket{\psi_{III}} \bra{\psi_{III}}
\end{equation}
where
$\ket{\psi_{III}} = a \ket{3} + b e^{i \delta} \ket{4}$ with $a^2 + b^2 = 1$. 
Since $\ket{\psi_{I}} = \openone \otimes \sigma_x \otimes \sigma_x \ket{\psi_{III}}$,  $(\openone \otimes \sigma_x \otimes \sigma_x) 
\hat{\rho}^T_{III} (0) (\openone \otimes \sigma_x \otimes \sigma_x)^{\dagger}$ has a GHZ-symmetry provided that $a^2 = b^2 = 1/2$ and 
$\delta = 0$.

Using Eqs. (\ref{diagonal-1}) and (\ref{non-diagonal-1}) one can show that the spectral decomposition of $\hat{\rho}^T_{III} (t)$ becomes
\begin{equation}
\label{type3-ghz-2}
\hat{\rho}^T_{III} (t) = \lambda_3 \ket{\phi_{III}} \bra{\phi_{III}} + (1 - P_t^2) \left[ a^2 (1 - P_t^2) + b^2  \right] \ket{0} \bra{0} + 
a^2 P_t^2 (1 - P_t^2) \left( \ket{1} \bra{1} + \ket{2} \bra{2} \right)
\end{equation}
where
\begin{eqnarray}
\label{type3-ghz-3}
&&\lambda_3 = P_t^2 (a^2 P_t^2 + b^2)                      \\    \nonumber
&& \ket{\phi_{III}} = \frac{1}{\sqrt{a^2 P_t^2+ b^2}} \left( a P_t \ket{3} + b  e^{i \delta} \ket{4} \right).
\end{eqnarray}
Unlike the case of type I $\hat{\rho}^T_{III} (t)$ is rank four tensor. From Eq. (\ref{type3-ghz-2}) one can derive the upper bound of $\tau_{ABC}$
for $\hat{\rho}^T_{III} (t)$, which is 
\begin{equation}
\label{upper-3}
\tau_{ABC} \leq \frac{4 a^2 b^2 P_t^4}{a^2 P_t^2 + b^2 }.
\end{equation}

The negativities  ${\cal N}_{A(BC)}$,  ${\cal N}_{B(AC)}$, and  ${\cal N}_{(AB)C}$ of  $\hat{\rho}^T_{III} (t)$ can be computed by making use of 
Eq. (\ref{negativity-1}), whose explicit expressions are 
\begin{eqnarray}
\label{ghz-3-pi-1}
&&{\cal N}_{A(BC)}  = \sqrt{(1 - P_t^2)^2 \left[a^2 (1 - P_t^2) + b^2 \right]^2 + 4 a^2 b^2 P_t^6} - (1 - P_t^2)  \left[a^2 (1 - P_t^2) + b^2 \right]
                                                                                                                                                                                        \nonumber   \\
&&{\cal N}_{B(AC)} = {\cal N}_{(AB)C} = \sqrt{a^4 P_t^4 (1 - P_t^2)^2 + 4 a^2 b^2 P_t^6} - a^2 P_t^2 (1 - P_t^2).
\end{eqnarray}
It is of interest to note that ${\cal N}_{A(BC)}$ and ${\cal N}_{B(AC)}$ of type III is the same with ${\cal N}_{(AB)C}$ and ${\cal N}_{A(BC)}$
of type II with $a \leftrightarrow b$ respectively.
It is easy to show ${\cal N}_{AB} = {\cal N}_{AC} = {\cal N}_{BC} = 0$. Thus the $\pi$-tangle of $\hat{\rho}^T_{III} (t)$ is 
\begin{equation}
\label{ghz-3-pi-2}
\pi_{GHZ}^{III} (t) = \frac{1}{3} \left[ {\cal N}_{A(BC)}^2 + 2 {\cal N}_{B(AC)}^2 \right].
\end{equation}

One can also consider different types of initial GHZ-type states. For example, one can consider 
$\hat{\rho}^T_{IV} (0) = \ket{\psi_{IV}} \bra{\psi_{IV}}$, where $\ket{\psi_{IV}} = a \ket{2} + b e^{i \delta} \ket{5}$. Although, in this case,
$\hat{\rho}^T_{IV} (t)$ is different from $\hat{\rho}^T_{II} (t)$, one can show that its $\pi$-tangle is exactly the same with that of type II. 
Thus, this case is not discussed in detail.

\begin{figure}[ht!]
\begin{center}
\includegraphics[height=5cm]{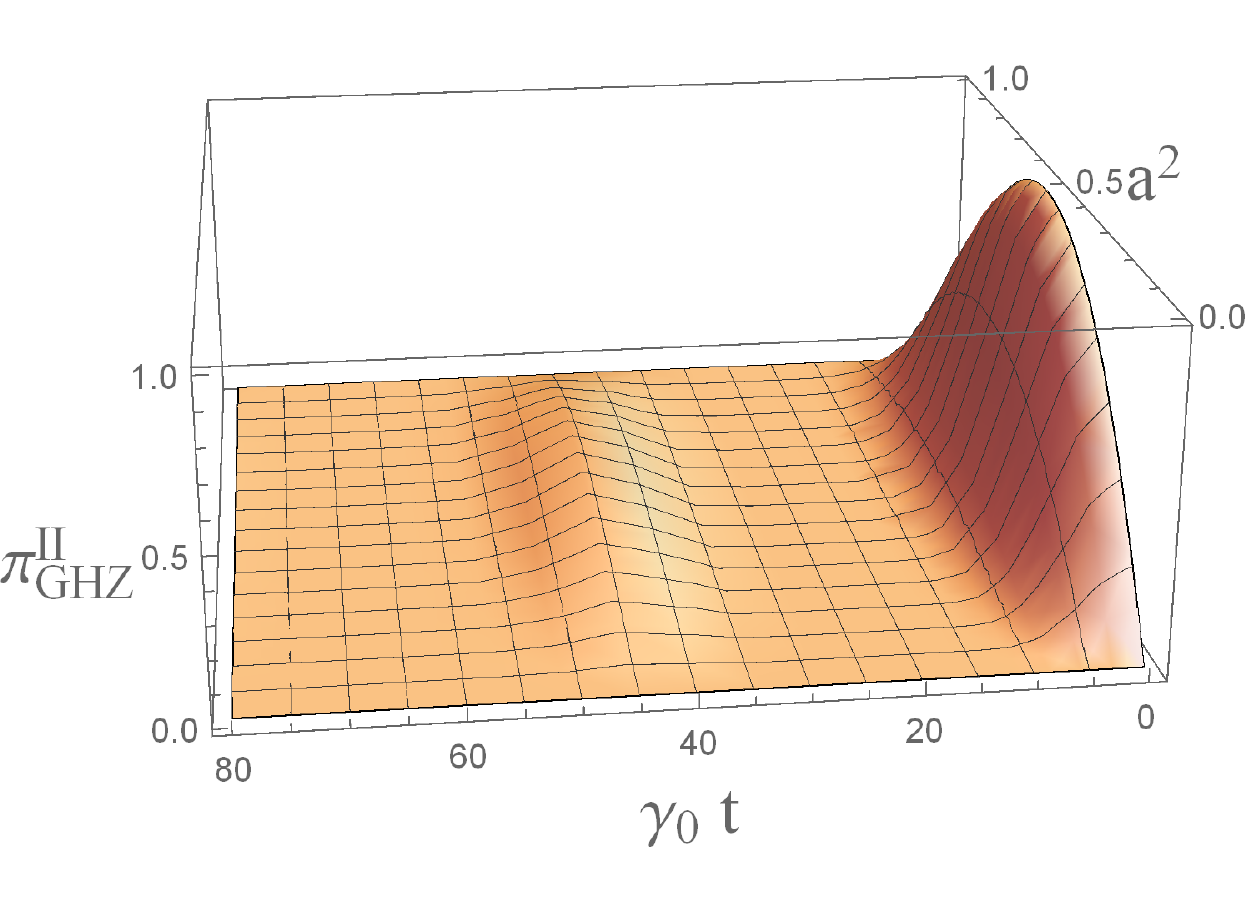}
\includegraphics[height=5cm]{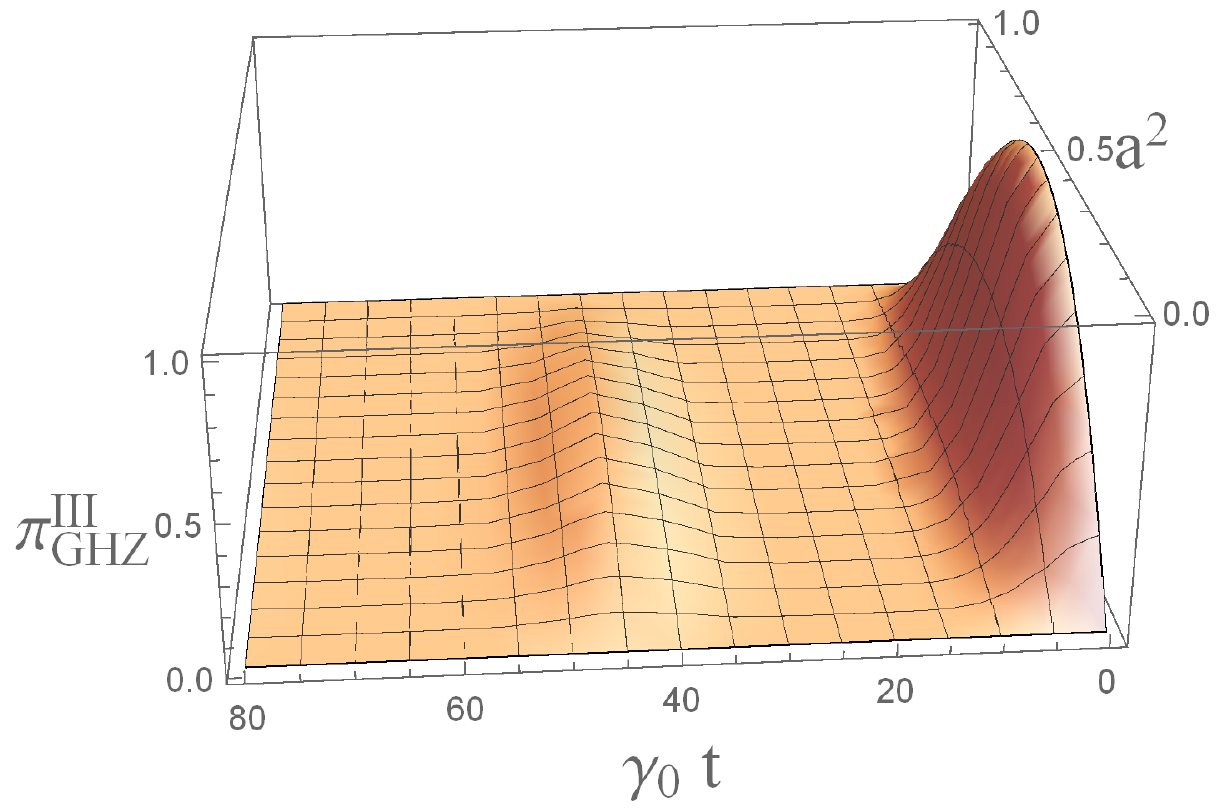}
\caption[fig2]{(Color online) The $\pi$-tangle for the initial states
(a) $ a \ket{001} + b e^{i \delta} \ket{110}$ and (b) $ a \ket{011} + b e^{i \delta} \ket{100}$ as a function of the parameters $\gamma_0 t$ 
and $a^2$. We choose $\lambda$ as a $\lambda = 0.01 \gamma_0$. }
\end{center}
\end{figure}

As shown in Eqs. (\ref{ghz-2-pi-2}) and (\ref{ghz-3-pi-2}) the dynamics of the tripartite entanglements for type II and type III are not expressed 
in terms of an inequality like Eq. (\ref{add1}) in type I. Thus, if $\ket{\psi_{II}}$ and $\ket{\psi_{III}}$ interact with the Markovian surroundings,
these entanglements decay exponentially with the half-life rule. This means that there is no ESD phenomenon in these cases. If  $\ket{\psi_{II}}$ 
and $\ket{\psi_{III}}$ interact with the non-Markovian environment, $\pi_{GHZ}^{II} (t)$ and $\pi_{GHZ}^{III} (t)$ should exhibit an oscillatory 
behavior with rapid decrease of amplitude due to $P_t$ in Eq. (\ref{pt-nm}). This can be seen in Fig. 2, where $\pi_{GHZ}^{II} (t)$ and 
$\pi_{GHZ}^{III} (t)$ are plotted as a function of dimensionless parameter $\gamma_0 t$ and $a^2$. We choose $\lambda$ as a
$\lambda = 0.01 \gamma_0$.  As expected the tripartite entanglement reduces to zero with increasing time with oscillatory behavior.


\begin{figure}[ht!]
\begin{center}
\includegraphics[height=5cm]{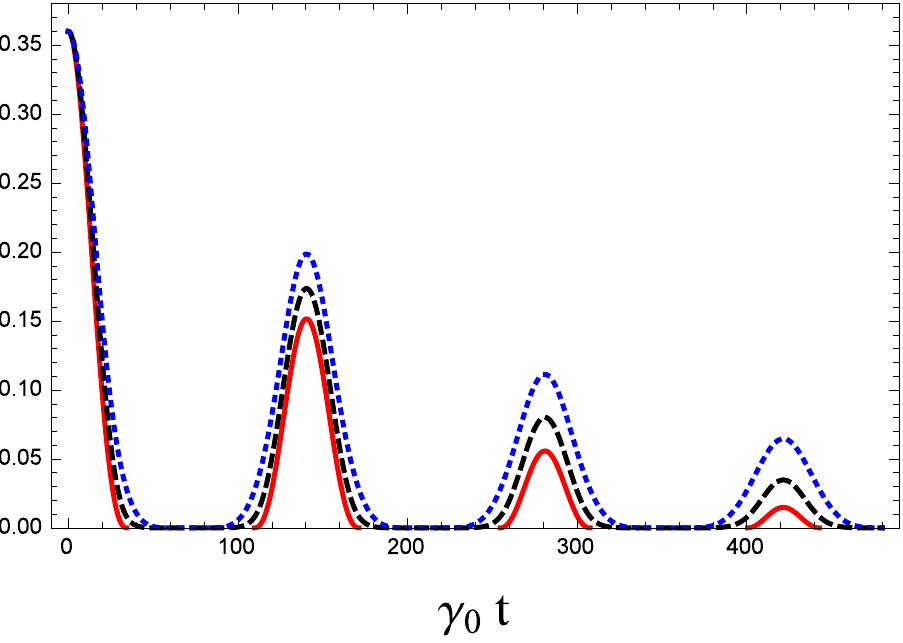}
\includegraphics[height=5cm]{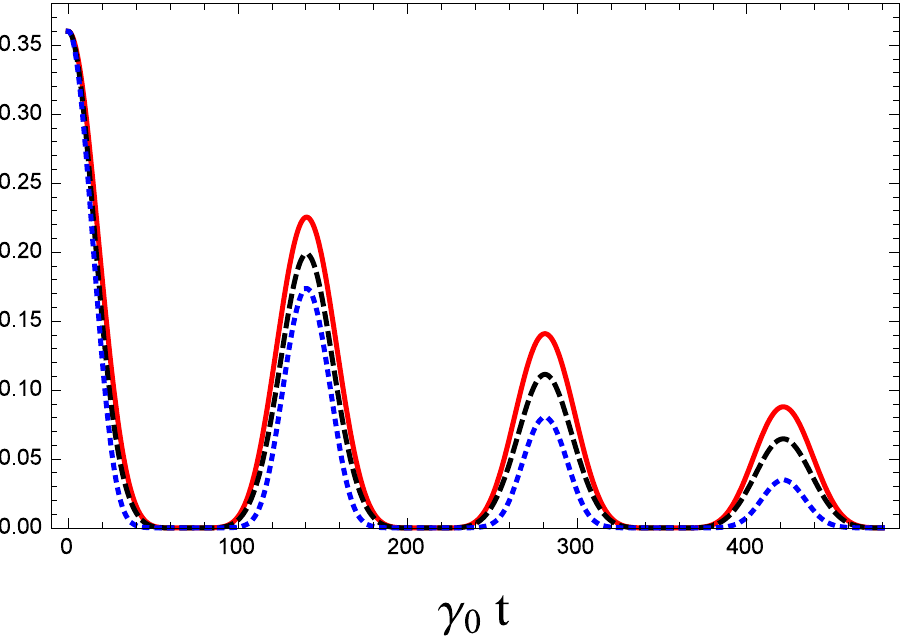}
\caption[fig3]{(Color online) The $\gamma_0 t$ dependence of $\pi_{GHZ}^{I} (t)$ (red solid), $\pi_{GHZ}^{II} (t)$ (black dashed), and 
$\pi_{GHZ}^{III} (t)$ (blue dotted) when (a) $a^2 = 0.1$ 
and (b) $a^2 = 0.9$. We choose $\lambda$ as a $\lambda = 0.001 \gamma_0$. }
\end{center}
\end{figure}

The $\pi$-tangles  $\pi_{GHZ}^{I} (t)$ , $\pi_{GHZ}^{II} (t)$ , and $\pi_{GHZ}^{III} (t)$ are compared in Fig. 3 when 
$\lambda / \gamma_0 = 0.001$. They are represented by red solid, black dashed, and blue dotted lines respectively. Fig. 3(a) and Fig. 3(b) 
correspond to $a^2 = 0.1$ and $a^2 = 0.9$. Both figures clearly show the revival of the tripartite entanglement, after a finite period of time of complete
disappearance. The revival phenomenon seems to be mainly due to the memory effect of the non-Markovian environment. It is of interest to note that
while  $\pi_{GHZ}^{III} (t)\geq  \pi_{GHZ}^{II} (t) \geq \pi_{GHZ}^{I} (t)$ when $a^2 = 0.1$, the order is changed as 
$\pi_{GHZ}^{I} (t)\geq  \pi_{GHZ}^{II} (t) \geq \pi_{GHZ}^{III} (t)$ when $a^2 = 0.9$.

\section{entanglement dynamics of W-type initial states}
In this section we examine the tripartite entanglement dynamics when the initial states are two W-type states. Both initial states are LU to 
each other. However, their entanglement dynamics are different due to Eqs. (\ref{diagonal-1}) and (\ref{non-diagonal-1}).

\subsection{Type I}
In this subsection we choose the initial state as 
\begin{equation}
\label{type1-W-1}
\hat{\rho}^W_I (0) = \ket{W_1} \bra{W_1}
\end{equation}
where $\ket{W_1} = a \ket{1} + b e^{i \delta_1} \ket{2} + c e^{i \delta_2} \ket{4}$ with $a^2 + b^2 + c^2 = 1$. Then, it is 
straightforward to show that the spectral decomposition of $\hat{\rho}^W_I (t)$ is 
\begin{equation}
\label{type1-W-2}
\hat{\rho}^W_I (t) = (1 - P_t^2) \ket{0} \bra{0} + P_t^2 \ket{W_1} \bra{W_1}.
\end{equation}
Eq. (\ref{type1-W-2}) guarantees that the residual entanglement and three-tangle of $\hat{\rho}^W_I (t)$ are zero because the spectral 
decomposition exactly coincides with the optimal decomposition.

By making use of Eq. (\ref{negativity-1}) one can compute the induced bipartite entanglement quantities
${\cal N}_{A(BC)}$,  ${\cal N}_{B(AC)}$, and  ${\cal N}_{(AB)C}$ 
of  $\hat{\rho}^W_{I} (t)$ directly, whose expressions are 
\begin{eqnarray}
\label{W-1-pi-1}
&& {\cal N}_{A(BC)} = \sqrt{(1 - P_t^2)^2 + 4 c^2 (a^2 + b^2) P_t^4} - (1 - P_t^2)       \nonumber   \\
&& {\cal N}_{B(AC)} = \sqrt{(1 - P_t^2)^2 + 4 b^2 (a^2 + c^2) P_t^4} - (1 - P_t^2)       \\      \nonumber
&& {\cal N}_{(AB)C} = \sqrt{(1 - P_t^2)^2 + 4 a^2 (b^2 + c^2) P_t^4} - (1 - P_t^2).
\end{eqnarray}
Also, the two tangles ${\cal N}_{AB}$,  ${\cal N}_{AC}$, and  ${\cal N}_{BC}$ become
\begin{eqnarray}
\label{W-1-pi-2}
&&{\cal N}_{AB} = \sqrt{\left[ (1 - P_t^2) + a^2 P_t^2 \right]^2 + 4 b^2 c^2 P_t^4} - \left[ (1 - P_t^2) + a^2 P_t^2 \right]   \nonumber  \\
&&{\cal N}_{AC} = \sqrt{\left[ (1 - P_t^2) + b^2 P_t^2 \right]^2 + 4 a^2 c^2 P_t^4} - \left[ (1 - P_t^2) + b^2 P_t^2 \right]    \\   \nonumber
&&{\cal N}_{BC} = \sqrt{\left[ (1 - P_t^2) + c^2 P_t^2 \right]^2 + 4 a^2 b^2 P_t^4} - \left[ (1 - P_t^2) + c^2 P_t^2 \right].
\end{eqnarray}

Thus, using Eqs. (\ref{pi-1}) and (\ref{pi-2}) one can compute the $\pi$-tangle of $\hat{\rho}^W_I (t)$, whose explicit expression is 
\begin{eqnarray}
\label{W-1-pi-3}
&&\pi^I_W (t) = \frac{2}{3} \Bigg[ 2 \left[(1 - P_t^2) + a^2 P_t^2 \right] \sqrt{\left[(1 - P_t^2) + a^2 P_t^2 \right]^2 + 4 b^2 c^2 P_t^4}  \nonumber \\
&&\hspace{2.0cm} + 2 \left[(1 - P_t^2) + b^2 P_t^2 \right] \sqrt{\left[(1 - P_t^2) + b^2 P_t^2 \right]^2 + 4 a^2 c^2 P_t^4}                   \nonumber   \\
&&\hspace{2.0cm} + 2 \left[(1 - P_t^2) + c^2 P_t^2 \right] \sqrt{\left[(1 - P_t^2) + c^2 P_t^2 \right]^2 + 4 a^2 b^2 P_t^4}                   \\  \nonumber
&&\hspace{2,0cm} - (1 - P_t^2) \bigg\{ \sqrt{(1 - P_t^2)^2 + 4 a^2 (b^2 + c^2) P_t^4}   \\   \nonumber
&&\hspace{2.0cm} + \sqrt{(1 - P_t^2)^2 + 4 b^2 (a^2 + c^2) P_t^4}                                   
+ \sqrt{(1 - P_t^2)^2 + 4 c^2 (a^2 + b^2) P_t^4} \bigg\}                                                          \\  \nonumber
&&\hspace{4.0cm} - 2 (a^4 + b^4 + c^4) P_t^4 - (1 - P_t^2) (3 + P_t^2)    \Bigg].
\end{eqnarray}
When $t = 0$, Eq. (\ref{W-1-pi-3}) reduces to 
\begin{equation}
\label{W-1-pi-4}
\pi^I_W (0) = \frac{4}{3} \left[a^2 \sqrt{a^4 + 4 b^2 c^2} + b^2 \sqrt{b^4 + 4 a^2 c^2} + c^2 \sqrt{c^4 + 4 a^2 b^2} 
- (a^4 + b^4 + c^4)  \right],
\end{equation}
which exactly coincides with a result of Ref.\cite{ou07-1}. Of course, when $t = t_n (n= 1, 2, \cdots)$ and $t = \infty$, the entanglement of 
$\hat{\rho}^W_I (t)$ is completely disentangled in the non-Markovian regime.

\subsection{Type II}
In this subsection we choose the initial state as 
\begin{equation}
\label{type2-W-1}
\hat{\rho}^W_{II} (0) = \ket{W_2} \bra{W_2}
\end{equation}
where $\ket{W_2} = a \ket{6} + b e^{i \delta_1} \ket{5} + c e^{i \delta_2} \ket{3}$ with $a^2 + b^2 + c^2 = 1$. 
This initial state is LU to $\ket{W_1}$ because of $\ket{W_2} = (\sigma_x \otimes \sigma_x \otimes \sigma_x) \ket{W_1}$.
Then, by making use of Eqs. (\ref{diagonal-1}) and (\ref{non-diagonal-1}) it is 
straightforward to show that  $\hat{\rho}^W_{II} (t)$ is 
\begin{equation}
\label{type2-W-2}
\hat{\rho}^W_{II} (t) = (1 - P_t^2)^2 \ket{0} \bra{0} + P_t^4 \ket{W_2} \bra{W_2} + 2 P_t^2 (1 - P_t^2) \sigma_{II} (t)
\end{equation}
where
\begin{eqnarray}
\label{type2-W-3}
&&\sigma_{II} (t) = \frac{1}{2} \Bigg[ (b^2 + c^2) \ket{1} \bra{1} + (a^2 + c^2) \ket{2} \bra{2} + (a^2 + b^2) \ket{4} \bra{4}   \nonumber  \\
&&\hspace{2.0cm}+ a b \left( e^{i \delta_1} \ket{1} \bra{2} + e^{-i \delta_1} \ket{2} \bra{1} \right) 
+ a c \left( e^{i \delta_2} \ket{1} \bra{4} + e^{-i \delta_2} \ket{4} \bra{1} \right)                                                     \\   \nonumber
&&\hspace{4.0cm}+ b c \left(e^{-i (\delta_1 - \delta_2)} \ket{2} \bra{4} + e^{i (\delta_1 - \delta_2)} \ket{4} \bra{2} \right) \Bigg].
\end{eqnarray}
The spectral decomposition of $\sigma_{II} (t)$ cannot be derived analytically. Also, analytic computation of $\pi$-tangle for 
$\hat{\rho}^W_{II} (t)$ is impossible. Thus, we have to reply on the numerical approach for computation of $\pi$-tangle. 

\begin{figure}[ht!]
\begin{center}
\includegraphics[height=5cm]{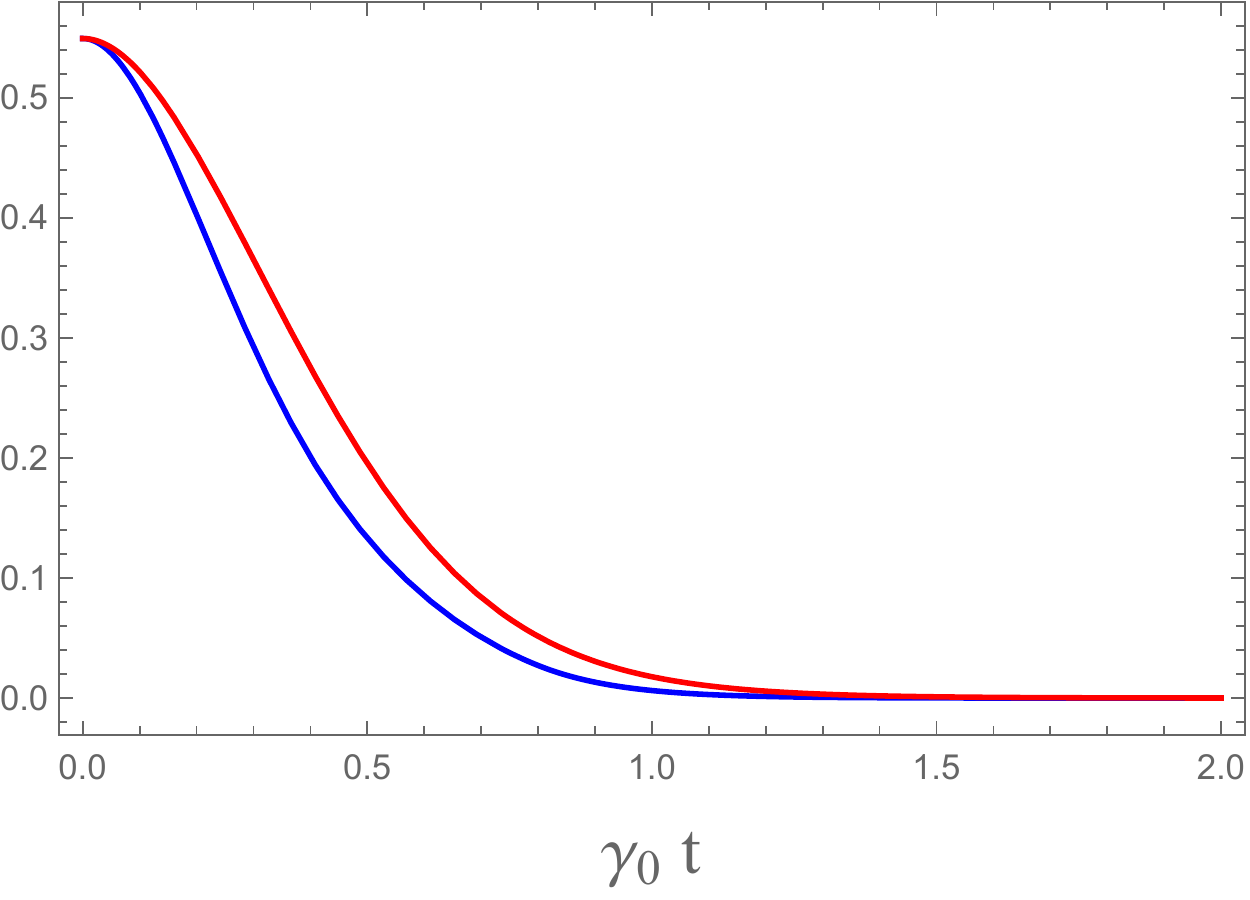}
\caption[fig4]{(Color online) The $\gamma_0 t$ dependence of $\pi^{I}_W$ (red line) and $\pi^{II}_W$ (blue line) 
when $\ket{W_1}$ and $\ket{W_2}$ interact with the Markovian environment. We choose $\lambda = 3 \gamma_0$ and $a^2 = b^2 = c^2 = 1/3$.}
\end{center}
\end{figure}

However, some special cases allow the analytic computation. In this paper we consider a special case $a^2 = b^2 = c^2 = 1/3$. In this case 
the spectral decomposition of $\sigma_{II} (t)$ can be derived as 
\begin{equation}
\label{type2-W-4}
\sigma_{II} (t) = \frac{2}{3} \ket{\alpha_1} \bra{\alpha_1} + \frac{1}{6}  \ket{\alpha_2} \bra{\alpha_2} + 
\frac{1}{6}  \ket{\alpha_3} \bra{\alpha_3}
\end{equation}
where
\begin{eqnarray}
\label{type2-W-5}
&&\ket{\alpha_1} = \frac{1}{\sqrt{3}} \left(\ket{1} + e^{-i \delta_1} \ket{2} + e^{-i \delta_2} \ket{4} \right)   \nonumber  \\
&&\ket{\alpha_2} = \frac{1}{\sqrt{2}} \left(\ket{1} - e^{-i \delta_2} \ket{4} \right)     \\    \nonumber
&&\ket{\alpha_3} = \frac{1}{\sqrt{6}} \left(\ket{1} - 2 e^{-i \delta_1} \ket{2} + e^{-i \delta_2} \ket{4} \right).
\end{eqnarray}
Thus, Eqs. (\ref{type2-W-2}) and (\ref{type2-W-4}) imply that $\hat{\rho}^W_{II} (t)$ with  $a^2 = b^2 = c^2 = 1/3$ is 
rank-$5$ tensor, three of them are W-states and the remaining ones are fully-separable and bi-separable states. Thus, its residual entanglement and 
three-tangles are zero.

Using Eq. (\ref{negativity-1}) one can show that  ${\cal N}_{A(BC)}$,  ${\cal N}_{B(AC)}$, and  ${\cal N}_{(AB)C}$ are all identical as 
\begin{equation}
\label{W-2-pi-1}
{\cal N}_{A(BC)} = {\cal N}_{B(AC)} = {\cal N}_{(AB)C} = \frac{1}{3} P_t^2 
\left[ \sqrt{9 - 18 P_t^2 + 17 P_t^4} - 3 (1 - P_t^2) \right].
\end{equation}
Also  ${\cal N}_{AB}$,  ${\cal N}_{AC}$, and  ${\cal N}_{BC}$ are all identical as 
\begin{eqnarray}
\label{W-2-pi-2}
 {\cal N}_{AB} = {\cal N}_{AC} = {\cal N}_{BC} = \left\{      \begin{array}{cc}
                                 \frac{\sqrt{9 - 24 P_t^2 + 20 P_t^4} + 2 P_t^2 (2 - P_t^2)}{3} - 1   & \hspace{1.0cm} P_t^2 \geq 2 - \sqrt{2}    \\
                                 0                & \hspace{1.0cm} P_t^2 \leq 2 - \sqrt{2}.
                                                                                                \end{array}            \right.   
\end{eqnarray} 
Thus, the $\pi$-tangle for $\hat{\rho}^W_{II} (t)$ with  $a^2 = b^2 = c^2 = 1/3$ is given by 
$\pi^{II}_W = {\cal N}_{A(BC)}^2 - 2 {\cal N}_{AB}^2$.

In Fig. 4 we plot $\pi^I_W (t)$ (red line) and $\pi^{II}_W (t)$ (blue line) as a function of $\gamma_0 t$ 
when $\ket{W_1}$ and $\ket{W_2}$ interact with the Markovian
environment. We choose $\lambda = 3 \gamma_0$ and $a^2 = b^2 = c^2 = 1/3$. As expected both reduce to zero with the half-life rule. It is of interest
to note $\pi^I_W (t) \geq \pi^{II}_W (t)$ in full range of time. This means that $\ket{W_1}$ is more robust than 
$\ket{W_2}$  against the Markovian environment.

\begin{figure}[ht!]
\begin{center}
\includegraphics[height=5cm]{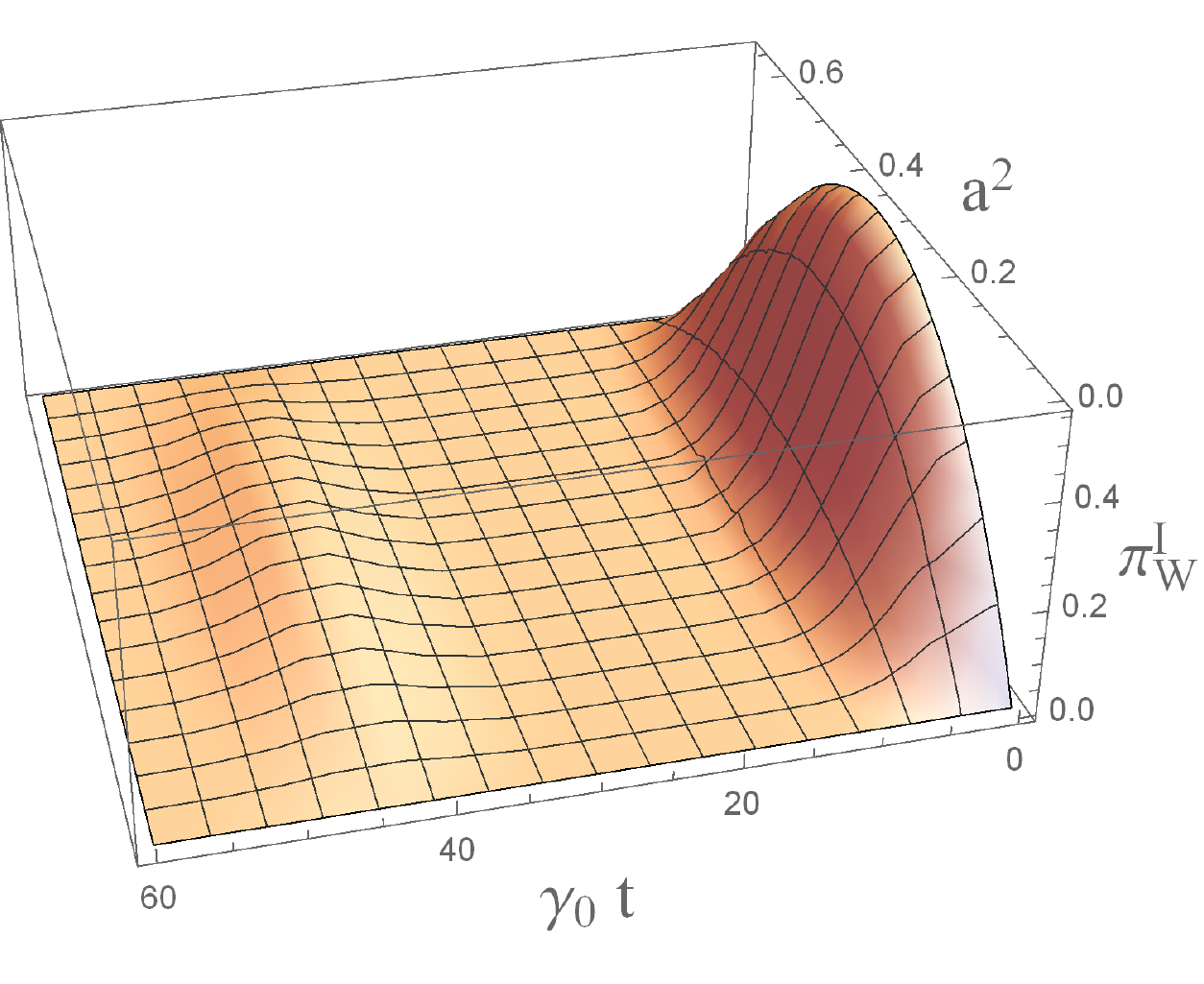}
\includegraphics[height=5cm]{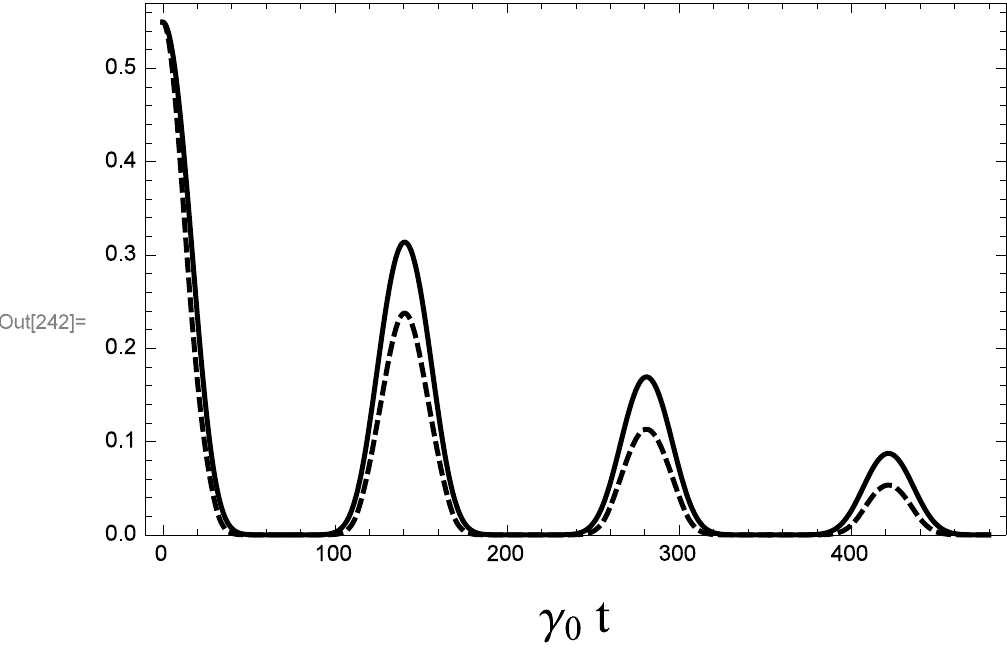}
\caption[fig5]{(Color online) (a) The 
$a^2$ and $\gamma_0 t$ dependence of $\pi_{W}^{I} (t)$ when $c^2 = 1/3$. We choose 
 $\lambda = 0.01 \gamma_0$.  (b) The  $\gamma_0 t$ dependence of  $\pi_{I}^{W} (t)$ (solid line) and  $\pi_{II}^{W} (t)$ (dashed line) when 
 $a^2 = b^2 = c^2 = 1/3$. We choose  $\lambda = 0.001 \gamma_0$. This figure implies that $\hat{\rho}^W_{I} (t)$ is more robust against the 
 environment than $\hat{\rho}^W_{II} (t)$. }
\end{center}
\end{figure}

In Fig. 5(a) we plot $\pi_{W}^{I} (t)$ as a function of $a^2$ and $\gamma_0 t$ when $\ket{W_1}$ is embedded  in the non-Markovian environment. 
We choose $c^2 = 1/3$ and $\lambda / \gamma_0 = 0.01$.
As expected the $\pi$-tangle reduces to zero as $t \rightarrow \infty$ with an oscillatory behavior. 
To compare  $\pi_{W}^{I} (t)$ with  $\pi_{W}^{II} (t)$ we plot 
both $\pi$-tangles as a function of $\gamma_0 t$ in Fig. 5(b). In this figure we choose $a^2 = b^2 = c^2 = 1/3$ and $\lambda / \gamma_0 = 0.001$.
The $\pi$-tangles  $\pi_{W}^{I} (t)$ and  $\pi_{W}^{II} (t)$ are plotted as solid and dashed lines respectively. In this case, as in the other 
cases, the revival of entanglement occurs after complete disappearance. It is interesting to note that like a Markovian case  
$\hat{\rho}^W_{I} (t)$ is more robust than $\hat{\rho}^W_{II} (t)$ against non-Markovian environment.

\section{Conclusions}

In this paper we have examined the tripartite entanglement dynamics when each party is entangled with other parties initially, 
but they locally interact with their 
own Markovian or non-Markovian environment. First, we have considered 
three GHZ-type initial states $\ket{\psi_I} = a \ket{000} + b e^{i \delta} \ket{111}$,
 $\ket{\psi_{II}} = a \ket{001} + b e^{i \delta} \ket{110}$, and  $\ket{\psi_{III}} = a \ket{011} + b e^{i \delta} \ket{100}$. 
 All states are LU to each other. 
  It turns out that the GHZ symmetry of the initial states is broken due to the effect of environment.
 We have computed the corresponding $\pi$-tangles analytically at arbitrary time $t$ in 
 Eqs. (\ref{ghz-1-pi-2}),  (\ref{ghz-2-pi-2}), and  (\ref{ghz-3-pi-2}).
 It was shown that while the ESD phenomenon occurs for type I, the entanglement dynamics for the remaining types exhibits an exponential 
 decay in the Markovian regime.
 In the non-Markovian regime the $\pi$-tangles completely vanish when 
 $t_n = 2 [n \pi - \tan^{-1} (d/\lambda) / d] \hspace{.3cm} (n = 1, 2, \cdots)$ and 
 $t \rightarrow \infty$. As shown in Fig. 3 the revival phenomenon of entanglement occurs after complete disappearance of entanglement. 
Based on the analytical results it was shown that while the robustness order against the effect of reservoir is 
 $\ket{\psi_I}$,  $\ket{\psi_{II}}$,  $\ket{\psi_{III}}$ for large $a^2$ region, this order is reversed for small $a^2$ region.
 
\begin{figure}[ht!]
\begin{center}
\includegraphics[height=5cm]{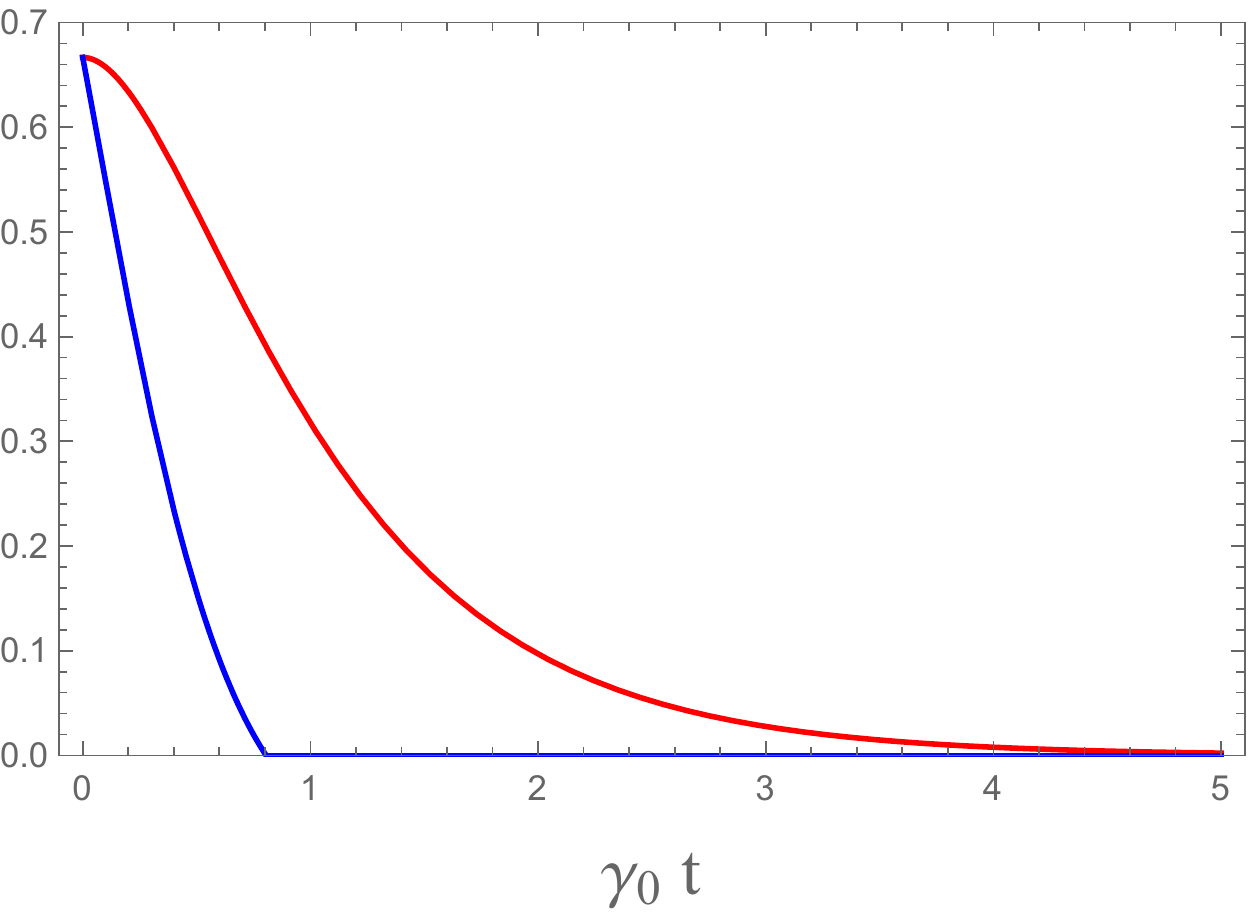}
\includegraphics[height=5cm]{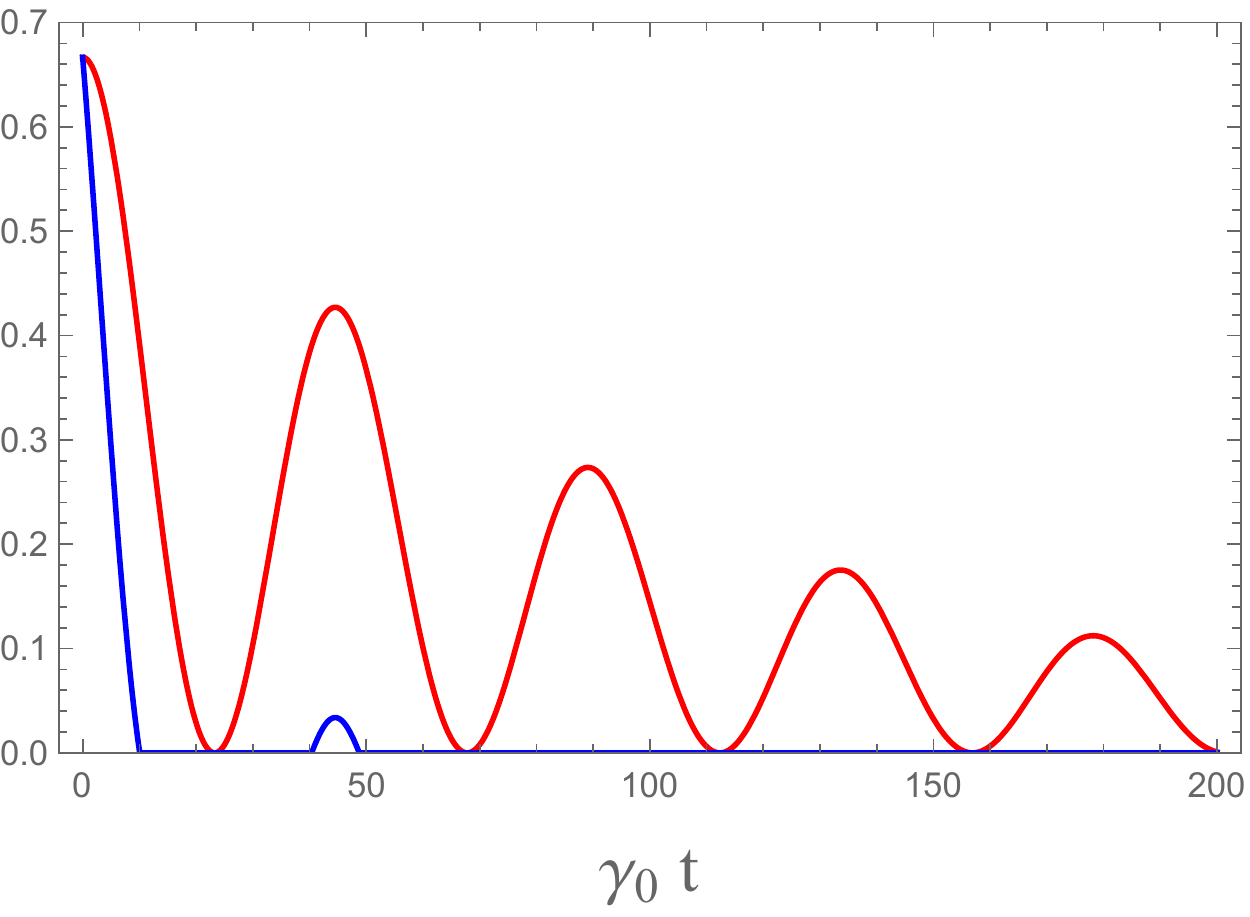}
\caption[fig5]{(Color online) The $\gamma_0 t$ dependence of concurrences Eq.(\ref{bipartite1}) and Eq. (\ref{bipartite2}) 
when $a^2 = b^2 = c^2 = 1/3$. (a) In this figure we choose $\lambda = 3 \gamma_0$. This shows that while bipartite entanglement dynamics 
for type I (red line) decays exponentially with the half-life rule, that for type II (blue line) exhibits an ESD. (b) In this figure we choose 
$\lambda = 0.01 \gamma_0$. Although both entanglements decay in time, the decay rate for type II (blue line) is much faster than that for type I (red line).}
\end{center}
\end{figure}
 
 We also have examined the tripartite entanglement dynamics for two W-type initial states 
 $\ket{W_1} = a \ket{001} + b e^{i \delta_1} \ket{010} + c e^{i \delta_2} \ket{100}$ and 
 $\ket{W_2} = a \ket{110} + b e^{i \delta_1} \ket{101} + c e^{i \delta_2} \ket{011}$ with $a^2 + b^2 + c^2 = 1$. 
 Like GHZ-type initial states they are LU to each other. For initial $\ket{W_1}$ state the $\pi$-tangle is analytically computed in Eq. (\ref{W-1-pi-3}).
 Since, however, $\ket{W_2}$ propagates to higher-rank state with the lapse of time, the analytic computation is impossible except few 
 special cases. Thus, we have computed the $\pi$-tangle analytically for special case $a^2 = b^2 = c^2 = 1/3$. In Fig. 4 and Fig. 5 it was shown that $\ket{W_1}$
 is more robust than $\ket{W_2}$ against the Markovian and non-Markovian environments. The bipartite entanglements measured by the 
 concurrence\cite{concurrence1} for $\hat{\rho}^W_I (t)$ and $\hat{\rho}^W_{II} (t)$ are
 \begin{equation}
 \label{bipartite1}
 {\cal C}^I_{AB} (t) = 2 |b c| P_t^2  \hspace{1.0cm} {\cal C}^I_{AC} (t) = 2 |a c| P_t^2  \hspace{1.0cm} {\cal C}^I_{BC} (t) = 2 |a b| P_t^2  
 \end{equation}
 and 
 \begin{eqnarray}
 \label{bipartite2}
 && {\cal C}^{II}_{AB} (t) = 2 P_t^2 \max \left[0, |b c| - |a| \sqrt{(1 - P_t^2) (1 - a^2 P_t^2)} \right]   \nonumber  \\
&& {\cal C}^{II}_{AC} (t) = 2 P_t^2 \max \left[0, |a c| -|b| \sqrt{(1 - P_t^2) (1 - b^2 P_t^2)} \right]    \\    \nonumber
&& {\cal C}^{II}_{BC} (t) = 2 P_t^2 \max \left[0, |a b| - |c| \sqrt{(1 - P_t^2) (1 - c^2 P_t^2)} \right].
 \end{eqnarray}
 One can show ${\cal C}^I \geq {\cal C}^{II}$ in the entire range of time like a tripartite entanglement regardless of Markovian or non-Markovian 
 environment. The $\gamma_0 t$-dependence of the concurrences is plotted in Fig. 6 as red line for type I and blue line for type II when 
 (a) Markovian ($\lambda = 3 \gamma_0$) and (b) non-Markovian ($\lambda = 0.01 \gamma_0$) environments are introduced. The Fig. 6(a) 
 shows that while the entanglement for type I exhibits an exponential decay with the half-life rule, that for type II exhibits an ESD. For non-Markovian case
 the decay rate for type II is much faster than that for type I although both exhibit a revival phenomenon of entanglement.
 
 It is of interest to study the effect of non-Markovian environment when the initial state is a rank-$2$ mixture
 \begin{equation}
 \label{conclusion-1}
 \rho(p) = p \ket{\mbox{GHZ}} \bra{\mbox{GHZ}} + (1 - p) \ket{\mbox{W}} \bra{\mbox{W}}
 \end{equation}
 where $ \ket{\mbox{GHZ}} = (\ket{000} + \ket{111}) / \sqrt{2}$ and $\ket{\mbox{W}} = (\ket{001} + \ket{010} + \ket{100}) / \sqrt{3}$.
 The residual entanglement of $\rho(p)$ is known as 
 \begin{eqnarray}
 \label{conclusion-2}
 \tau(p) = \left\{                           \begin{array}{cc}
                           0  &  \hspace{1.0cm} 0 \leq p \leq p_0                                \\
                           g_I(p)  &  \hspace{1.0cm}  p_0 \leq p \leq p_1                      \\
                           g_{II} (p)  &   \hspace{1.0cm} p_1 \leq p \leq 1
                                                       \end{array}                      \right.
\end{eqnarray}
where 
\begin{eqnarray}
\label{conclusion-3}
&&p_0 = \frac{4 \sqrt[3]{2}}{3 + 4 \sqrt[3]{2}} = 0.626851\cdots   
\hspace{1.0cm} p_1 = \frac{1}{2} + \frac{3 \sqrt{465}}{310} = 0.70868\cdots                      \\   \nonumber
&&g_I (p) = p^2 - \frac{8 \sqrt{6}}{9} \sqrt{p (1 - p)^3}                  \hspace{1.0cm}
g_{II}(p) = 1 - (1 - p) \left(\frac{3}{2} + \frac{1}{18} \sqrt{465} \right).
\end{eqnarray}
It is interesting, at least for us, how the non-Markovian environment modifies Coffman-Kundu-Wootters inequality 
$4 \min [\mbox{det} (\rho_A)] \geq {\cal C}(\rho_{AB})^2 +  {\cal C}(\rho_{AC})^2$ in this model. Similar issue was discussed in 
Ref. \cite{costa14}.

Since we have derived the $\pi$-tangles analytically, we tried to find the entanglement invariants\cite{yu09-1,yonac07}, which was originally found in 
four-qubit system. In our three-qubit systems we cannot find any invariants. It is of interest to examine the entanglement invariants in the 
higher-qubit and qudit systems.





\begin{thebibliography}{99}
\bibitem{nielsen00} M. A. Nielsen and I. L. Chuang, {\it Quantum Computation and
Quantum Information} (Cambridge University Press, Cambridge, England, 2000).
\bibitem{horodecki09} R. Horodecki, P. Horodecki, M. Horodecki, and K. Horodecki, {\it Quantum Entanglement}, Rev. Mod. Phys. 
{\bf 81} (2009) 865 [quant-ph/0702225] and references therein.
\bibitem{teleportation} C. H. Bennett, G. Brassard, C. Cr´epeau, R. Jozsa, A. Peres, and W. K. Wootters, {\it Teleporting
an Unknown Quantum State via Dual Classical and Einstein-Podolsky-Rosen Channles}, Phys.Rev. Lett. {\bf 70} (1993) 1895.
\bibitem{superdense} C. H. Bennett and S. J. Wiesner, {\it Communication via one- and two-particle operators on
Einstein-Podolsky-Rosen states}, Phys. Rev. Lett. {\bf 69} (1992) 2881.
\bibitem{clon} V. Scarani, S. Lblisdir, N. Gisin, and A. Acin, {\it Quantum cloning}, Rev. Mod. Phys. {\bf 77} (2005)
1225 [quant-ph/0511088] and references therein.
\bibitem{cryptography} A. K. Ekert , {\it Quantum Cryptography Based on Bell’s Theorem}, Phys. Rev. Lett. {\bf 67} (1991)
661.
\bibitem{cryptography2} C. Kollmitzer and M. Pivk, Applied Quantum Cryptography (Springer, Heidelberg, Germany, 2010).
\bibitem{qcreview} T. D. Ladd, F. Jelezko, R. Laflamme, Y. Nakamura, C. Monroe, and J. L. O'Brien, 
{\it Quantum Computers}, Nature, {\bf 464} (2010) 45 [arXiv:1009.2267 (quant-ph)].
\bibitem{qcomputer} G. Vidal, {\it Efficient classical simulation of slightly entangled quantum computations}, Phys. Rev.
Lett. {\bf 91} (2003) 147902 [quant-ph/0301063].
\bibitem{breuer02} H. -P. Breuer and F. Petruccione, {\it The Theory of Open Quantum Systems} (Oxford University Press, 
Oxford, New York, 2002). 
\bibitem{zurek03} W. H. Zurek, {\it Decoherence, einselection, and the quantum origins of the classical}, Rev. Mod. Phys. {\bf 75} (2003) 715
[quant-ph/0105127].
\bibitem{yu02-1} T. Yu and J. H. Eberly, {\it Phonon decoherence of quantum entanglement: Robust and fragile states}, Phys. Rev. {\bf B 66}
(2002) 193306 [quant-ph/0209037].
\bibitem{simon02-1} C. Simon and J. Kempe, {\it Robustness of multiparty entanglement}, Phys. Rev. {\bf A 65} (2002) 052327 [quant-ph/0109102].
\bibitem{dur04-1} W. D\"{u}r and H. J. Briegel, {\it Stability of Macroscopic Entanglement under Decoherence}, Phys. Rev. Lett. {\bf 92} (2004)
180403 [quant-ph/0307180].
\bibitem{markovian} T. Yu and J. H. Eberly, {\it Finite-Time Disentanglement Via Spontaneous Emission}, Phys. Rev. Lett. {\bf 93}
(2004) 140404 [quant-ph/0404161].
\bibitem{yu05-1}  T. Yu and J. H. Eberly, {\it Sudden Death of Entanglement: Classical Noise Effects}, Opt. Commun. {\bf 264} (2006) 393
[quant-ph/0602196].
\bibitem{yu06-1} T. Yu and J. H. Eberly, {\it Quantum Open System Theory: Bipartite Aspects}. Phys. Rev. Lett. {\bf 97} (2006) 140403
[quant-ph/0603256]
\bibitem{yu09-1} T. Yu and J. H. Eberly, {\it Sudden Death of Entanglement}, Science, {\bf 323} (2009) 598 [arXiv:0910.1396 (quant-ph)].
\bibitem{almeida07} M.P. Almeida {\it et al}, {\it Environment-induced Sudden Death of Entanglement}, Science {\bf 316}
(2007) 579 [quant-ph/0701184].
\bibitem{laurat07} J. Laurat, K. S. Choi, H. Deng, C. W. Chou, and H. J. Kimble, {\it Heralded Entanglement between Atomic Ensembles: Preparation, Decoherence, and Scaling}, Physics. Rev. Lett. {\bf 99} (2007) 180504 [arXiv:0706.0528 (quant-ph)].
\bibitem{lopez08} C. E. L\'opez, G. Romero, F. Lastra, E. Solano, and J. C. Retamal, 
{\it Sudden Birth versus Sudden Death of Entanglement in Multipartite Systems}, Phys. Rev. Lett. {\bf 101} (2008) 080503 [arXiv:0802.1825 (quant-ph)].
\bibitem{bellomo07}B. Bellomo, R. Lo Franco, and G. Compagno, {\it Non-Markovian Effects on the Dynamics of Entanglement}, Phys. Rev. Lett. {\bf 99}
(2007) 160502 [arXiv:0804.2377 (quant-ph)].
\bibitem{concurrence1} S. Hill and W. K.  Wootters, {\it Entanglement of a pair of quantum bits}, Phys. Rev. Lett. {\bf 78} (1997) 5022 
[quant-ph/9703041; W. K. Wootters, {\it Entanglement of Formation of an Arbitrary State of
Two Qubits}, Phys. Rev. Lett. {\bf80} (1998) 2245 [quant-ph/9709029].
\bibitem{breuer09}H. -P.  Breuer,  E. -M. Laine, and J. Piilo, {\it Measure for the Degree of Non-Markovian Behavior of Quantum Processes in Open Systems}, Phys. Rev. Lett. {\bf 103} (2009) 210401 [arXiv:0908.0238 (quant-ph)].
\bibitem{vacchini11} B. Vacchini, A. Smirne, E. -M. Laine, J. Piilo, and H. -P. Breuer, {\it Markovian and non-Markovian dynamics in quantum and classical systems},  New J. Phys. {\bf 13} (2011) 093004 [arXiv:1106.0138 (quant-ph)].
\bibitem{chruscinski11} D. Chru\'sci\'nski, A. Kossakowski, and A. Rivas, {\it Measures of non-Markovianity: Divisibility versus backflow of information}, 
Phys. Rev. {\bf A 83}  (2011) 052128  [arXiv:1102.4318 (quant-ph)].
\bibitem{rivas14} A. Rivas, S. F. Huelga, and M. B. Plenio, {\it Quantum Non-Markovianity: characterization, quantification and detection},
Rep. Prog. Phys. {\bf 77}  (2014) 094001 [arXiv:1405.0303 (quant-ph)].
\bibitem{hall14} M. J. W. Hall, J. D. Cresser, L. Li, and E. Andersson {\it Canonical form of master equations and characterization of non-Markovianity}, 
Phys. Rev. {\bf A 89} (2014) 042120 [arXiv:1009.0845 (quant-ph)].
\bibitem{kwang15-1} K .-I. Kim, H .-M. Li, and B. -K. Zhao, 
{\it GenuineTripartite Entanglement Dynamics and Transfer in a Triple Jaynes-Cummings Model}, Int. J. Theor. Phys. {\bf 55} (2016) 241.
\bibitem{yonac07} M. Y\"{o}nac, T. Yu, and J. H. Eberly, {\it Pairwise concurrence dynamics: a four-qubit model}, J. Phys. B: At. Mol. Opt. Phys. 
{\bf 40} (2007) 545 [quant-ph/0701111]. 
\bibitem{green89} D. M. Greenberger, M. Horne, and A. Zeilinger, {\it Bell's Theorem,
Quantum Theory, and Conceptions of the Universe}, edited by M. Kafatos (Kluwer,
Dordrecht, 1989).
\bibitem{dur00-1} W. D\"{u}r, G. Vidal, and J. I. Cirac, {\it Three qubits can be
entangled in two inequivalent ways}, Phys. Rev. {\bf A62} (2000) 062314
[quant-ph/0005115].
\bibitem{ckw} V. Coffman, J. Kundu, and W. K. Wootters, {\it Distributed entanglement}, Phys. Rev. {\bf A 61} (2000) 052306 [quant-ph/9907047].
\bibitem{ou07-1} Y. U. Ou and H. Fan, {\it Monogamy Inequality in terms of Negativity for
Three-Qubit States}, Phys. Rev. {\bf A75} (2007) 062308 [quant-ph/0702127].
\bibitem{benn96} C. H. Bennett, D. P. DiVincenzo, J. A. Smokin, and W. K. Wootters,
{\it Mixed-state entanglement and quantum error correction}, Phys. Rev. {\bf A 54} (1996) 3824 [quant-ph/9604024].
\bibitem{uhlmann99-1} A. Uhlmann, {\it Fidelity and concurrence of conjugate states},
Phys. Rev. {\bf A 62} (2000) 032307 [quant-ph/9909060].
\bibitem{residual} R. Lohmayer, A. Osterloh, J. Siewert, and A. Uhlmann, {\it Entangled
Three-Qubit States without Concurrence and Three-Tangle}, Phys. Rev. Lett. {\bf 97}
(2006) 260502 [quant-ph/0606071];
C. Eltschka, A. Osterloh, J. Siewert, and A. Uhlmann, {\it Three-tangle
for mixtures of generalized GHZ and generalized W states}, New J. Phys. {\bf 10} (2008)
043014 [arXiv:0711.4477 (quant-ph)];
E. Jung, M. R. Hwang, D. K. Park, and J. W. Son, {\it Three-tangle
for Rank-$3$ Mixed States: Mixture of Greenberger-Horne-Zeilinger, W and flipped W states},
Phys. Rev. {\bf A 79} (2009) 024306 [arXiv:0810.5403 (quant-ph)];
E. Jung, D. K. Park, and J. W. Son, {\it Three-tangle does not properly
quantify tripartite entanglement for Greenberger-Horne-Zeilinger-type state},
Phys. Rev. {\bf A 80} (2009) 010301(R) [arXiv:0901.2620 (quant-ph)];
E. Jung, M. R. Hwang, D. K. Park, and S. Tamaryan, {\it Three-Party Entanglement in Tripartite Teleportation
Scheme through Noisy Channels}, Quant. Inf. Comp. {\bf 10} (2010) 0377 [arXiv:0904.2807 (quant-ph)].
\bibitem{elts12-1} C. Eltschka and J. Siewert, {\it Entanglement of Three-Qubit Greenberger-Horne-Zeilinger-Symmetric States}, 
Phys. Rev. Lett. {\bf 108} (2012) 020502 [ arXiv:1304.6095 (quant-ph)].
\bibitem{siewert12-1} J. Siewert and C. Eltschka, {\it Quantifying Tripartite Entanglement of Three-Qubit Generalized Werner States}, 
Phys. Rev. Lett. {\bf 108} (2012) 230502.
\bibitem{vidal01-1} G. Vidal and R. F. Werner, {\it Computable measure of entanglement},
Phys. Rev. {\bf A65} (2002) 032314 [quant-ph/0102117].
\bibitem{garraway97} B. M. Garraway, {\it Nonperturbative decay of an atomic system in a cavity}, Phys. Rev. {\bf A55} (1997) 2290.
\bibitem{manis06} S. Maniscalco and F. Petruccione, {\it Non-Markovian dynamics of a qubit}, Phys. Rev. {\bf A73} (2006) 
012111 [quant-ph/0509208].
\bibitem{kraus83} K. Kraus, {\it States, Effect, and Operations: Fundamental Notions in Quantum Theory} (Springer-Verlag, Berlin, 1983).
\bibitem{costa14} A. C. S. Costa, R. M. Angelo, and M. W. Beims, {\it Monogamy and backflow of mutual information in non-Markovian thermal baths},
Phys. Rev. {\bf A 90} (2014) 012322 [arXiv:1404.6433 (quant-ph)].






 \end{thebibliography}
\end{document}